\documentclass[preprint,12pt]{aastex}

\newcommand{\ee}[1]{\mbox{${} \times 10^{#1}$}}

\newcommand{\am}{\mbox{\arcmin}}
\newcommand{\as}{\mbox{\arcsec}}

\newcommand{\um}{$\mu$m}

\newcommand{\lsun}{\mbox{L$_\odot$}}
\newcommand{\msun}{\mbox{M$_\odot$}}
\newcommand{\rsun}{\mbox{R$_\odot$}}

\newcommand{\lbol}{\mbox{$L_{bol}$}} 
\newcommand{\lobs}{\mbox{$L_{obs}$}} 
\newcommand{\lint}{\mbox{$L_{int}$}} 
\newcommand{\tbol}{\mbox{$T_{bol}$}} 

\newcommand{\lobssmm}{$\lobs/$\lsmm}

\newcommand{\lsmm}{\mbox{$L_{smm}$}} 

\newcommand{\nthp}{\mbox{{\rm N$_2$H}$^+$}}

\newcommand{\andre}{Andr\'{e}}
\newcommand{\duchene}{Duch\^{e}ne}
\begin{document}

\title {The Spitzer c2d Survey of Nearby Dense Cores: I: First Direct Detection of the Embedded Source in IRAM 04191+1522}

\author{
Michael M. Dunham\altaffilmark{1,2},
Neal J. Evans II\altaffilmark{1},
Tyler L. Bourke\altaffilmark{3},
Cornelis P. Dullemond\altaffilmark{4},
Chadwick H. Young\altaffilmark{5},
Timothy Y. Brooke\altaffilmark{6},
Nicholas Chapman\altaffilmark{7},
Philip C. Myers\altaffilmark{3},
Alicia Porras\altaffilmark{3,8}
William Spiesman\altaffilmark{1},
Peter J. Teuben\altaffilmark{7},
Zahed Wahhaj\altaffilmark{9}}

\altaffiltext{1}{Department of Astronomy, The University of Texas at Austin, 1 University Station, C1400, Austin, Texas 78712--0259}

\altaffiltext{2}{E-mail: mdunham@astro.as.utexas.edu}

\altaffiltext{3}{Harvard-Smithsonian Center for Astrophysics, 60 Garden Street, Cambridge, MA 02138}

\altaffiltext{4}{Max-Planck-Institut fur Astronomie, Koenigstuhl 17, 69117 Heidelberg}

\altaffiltext{5}{Department of Physical Sciences, Nicholls State University, Thibodaux, LA 70301}

\altaffiltext{6}{Astronomy Department, MC 105-24, California Institute of Technology, Pasadena, CA 91125}

\altaffiltext{7}{Department of Astronomy, University of Maryland, College Park, MD 20742}

\altaffiltext{8}{Instituto Nacional de Astrof\'{\i}sica, \'Optica y
Electr\'onica (INAOE), Tonantzintla, Pue., M\'exico}

\altaffiltext{9}{Department of Physics and Astronomy, Northern Arizona University, Box 6010, Flagstaff, AZ 86011-6010}

\begin{abstract}
We report the first detections of the Class 0 protostellar source IRAM 04191+1522 at wavelengths shortward of 60 \um\ with the \emph{Spitzer Space Telescope}.  We see extended emission in the \emph{Spitzer} images that suggests the presence of an outflow cavity in the circumstellar envelope.  We combine the \emph{Spitzer} observations with existing data to form a complete dataset ranging from 3.6 to 1300 \um\ and use these data to construct radiative transfer models of the source.  We conclude that the internal luminosity of IRAM 04191+1522, defined to be the sum of the luminosity from the internal sources (a star and a disk), is $L_{int} = 0.08 \pm 0.04$ $\lsun$, placing it among the lowest luminosity protostars known.  Though it was discovered before the launch of the \emph{Spitzer Space Telescope}, IRAM 04191+1522 falls within a new class of Very Low Luminosity Objects being discovered by \emph{Spitzer}.  Unlike the two other well-studied objects in this class, which are associated either with weak, compact outflows or no outflows at all, IRAM 04191+1522 has a well-defined molecular outflow with properties consistent with those expected based on relations derived from higher luminosity ($L_{int} \geq 1$ \lsun) protostars.  We discuss the difficulties in understanding IRAM 04191+1522 in the context of the standard model of star formation, and suggest a possible explanation for the very low luminosity of this source.
\end{abstract}



\section{Introduction}\label{intro}

Despite substantial study and progress in recent decades, the initial conditions of isolated low-mass star formation remain poorly understood.  In the standard picture (Shu 1977; Shu, Adams, \& Lizano 1987), collapse begins from an envelope initially described by a singular isothermal sphere at rest with an initial density distribution of $\rho (r)\propto r^{-2}$.  Observations of star-forming cores in the submillimeter and millimeter wavelength regimes have generally supported the presence of power-law density distributions (e.g., Shirley et al. 2002; Motte \& \andre\ 2001).  However, observations of starless cores, defined to be cores without embedded sources detected by the \emph{Infrared Astronomical Satellite (IRAS)}, have shown that they are often not consistent with power-law density distributions, and instead show flat density profiles in their inner regions (e.g., Ward-Thompson et al. 1994; Evans et al. 2001), reminiscent of Bonnor-Ebert spheres (Bonnor 1956; Ebert 1955).  To explain this difference in physical conditions between starless cores and star-forming cores, recent work has concentrated on developing an evolutionary picture for starless cores and identifying a subset of ``evolved'' starless cores with high central densities, molecular depletion, and infall asymmetry that suggest they are close to the onset of collapse and star formation (e.g., Crapsi et al. 2005a; Kirk et al. 2005).

Recently, the \emph{Spitzer Space Telescope} Legacy Project ``From Molecular Cores to Planet Forming Disks'' (c2d; Evans et al. 2003) has surveyed 84 regions with isolated, dense cores, of which approximately 60 were classified as starless.  The very first of these starless cores observed by \emph{Spitzer}, L1014, was found to contain a very low luminosity object ($L \sim 0.09$ $\lsun$; Young et al. 2004a).  However, L1014 does not show evidence of being an evolved core that is close to star formation (Crapsi et al. 2005a; Crapsi et al. 2005b), suggesting that our understanding of the progression from starless core to protostar is not yet complete.

Other low luminosity objects are being discovered with \emph{Spitzer} due to its very high sensitivity in the mid-infrared.  This has given rise to a new class of objects called Very Low Luminosity Objects (VeLLOs; Di Francesco et al. 2006; Huard et al. in preparation).  Defining the internal luminosity of a source, \lint, to be the total luminosity of the central protostar and circumstellar disk (if present), a VeLLO is defined to be an object embedded within a dense core that meets the criterion $\lint \leq 0.1$ \lsun.  Assuming spherical mass accretion at the rate predicted by the standard model ($\dot{M}_{acc} \sim 2 \ee{-6}$ \msun\ yr$^{-1}$; Shu, Adams, \& Lizano 1987) onto an object with a typical protostellar radius of $R \sim 3$ \rsun, a protostar located on the stellar/substellar boundary ($M=0.08$ \msun) would have an accretion luminosity, $L_{acc} = \frac{GM\dot{M}_{acc}}{R}$, of $L \sim 1.6$ \lsun.  VeLLOs, with luminosities more than an order of magnitude lower than this, are difficult to understand in the context of the standard model of star formation.  If these objects were observed edge-on through circumstellar disks their luminosities could possibly be underestimated, but, for at least some of them, this possibility can be eliminated (see discussion in \S \ref{lint}).  Thus, they must either feature mass accretion rates much lower than predicted by the standard model, masses far below the stellar/substellar boundary, or some combination of the two.  These objects may pose both a challenge to our understanding of star formation and an opportunity to observe embedded proto-brown dwarfs.  It is important to examine them in detail in order to determine their physical properties and ultimately their place in the formation of low-mass stars and brown dwarfs.

This paper presents new mid-infrared observations of IRAM 04191+1522 (hereafter IRAM 04191), a young, low luminosity, Class 0 source in the Taurus region, along with models constructed to derive the properties of this object.  In \S \ref{04191info} we discuss previous work on IRAM 04191.  \S \ref{observations} details the observational data used in this paper, while \S \ref{results} presents images and photometry, along with a qualitative discussion of the results from these new data.  In \S \ref{models} we present one- and two-dimensional models that attempt to reproduce the observed Spectral Energy Distribution (SED) and gain insight into the luminosity of this source, and \S \ref{vello} discusses IRAM 04191 in the context of VeLLOs.  Finally, we present our conclusions in \S \ref{conclusions}.

\section{IRAM 04191+1522}\label{04191info}

IRAM 04191 is a Class 0 source located in the southern part of the Taurus molecular cloud at a distance of 140 pc (Kenyon et al. 1994), approximately 1\am\ to the southwest of the Class I source IRAS 04191+1523 (hereafter IRAS 04191)\footnote{Class 0 objects, indistinguishable from Class I objects in the Lada (1987) definition based on the slope of the infrared SED, are defined conceptually as protostars with envelope masses still exceeding the central protostellar mass, and observationally as objects with bolometric-to-submillimeter luminosity ratios of less than 200 (\andre, Ward-Thompson, \& Barsony 1993).}.  It was discovered during 1.3 mm dust continuum observations of Taurus (\andre\ et al. 1999).  \andre\ et al. presented observations ranging from 7.75 to 1300 \um\ that resulted in detections longward of 60 \um\ and non-detections shortward of 60 \um\ (see their Figure 3).  Despite the fact that the shortest wavelength at which IRAM 04191 was detected was 60 \um, the presence of an embedded protostar was inferred based on both excess emission in the $60-90$ \um\ range over that of thermal emission from cold dust heated only by the Interstellar Radiation Field (ISRF), and the presence of a molecular outflow (see below).  Based on the detections longward of 60 \um, \andre\ et al. calculated an observed bolometric luminosity (see \S \ref{results}) and temperature\footnote{The bolometric temperature of a source, \tbol, is defined to be the temperature of a blackbody with the same flux-weighted mean frequency as the source.} of \lobs\ $\sim$ 0.15 \lsun\ and \tbol\ $\sim$ 18 K, making IRAM 04191 one of the lowest luminosity protostars then known.  Since \andre\ et al.'s $\lobs \sim 0.15$ \lsun\ result includes a component from the ISRF, which can provide up to a few tenths of a solar luminosity to the bolometric luminosity (e.g., Shirley et al. 2000; Evans et al. 2001), IRAM 04191 may have a low enough internal luminosity to qualify as a VeLLO.  This claim and its implications are investigated in this work.

A well-defined CO (2-1) bipolar molecular outflow has been detected (\andre\ et al. 1999) and associated with IRAM 04191.  It extends from the northeast (redshifted) to the southwest (blueshifted).  Based on the fact that the outflow shows a clear bipolar morphology with almost no overlap between blueshifted and redshifted emission, along with the observed aspect ratio of $\sim$ 0.65 for the core, as observed in both dust and N$_2$H$^+$ emission, \andre\ et al. estimate the inclination to be $i \sim 50\mbox{$^{\circ}$}$ (where $i = 0\mbox{$^{\circ}$}$ corresponds to a pole-on orientation).  Other molecular line observations of this source have been obtained (Belloche et al. 2002; Takakuwa et al. 2003; Belloche \& \andre\ 2004; Lee et al. 2005) and indicate the presence of infall and rotation.  Based on its position in \lbol\ - \tbol\ evolutionary diagrams, \andre\ et al. (1999) estimate an age of approximately $1-3$ \ee{4} years for IRAM 04191.  Although this age is uncertain, IRAM 04191 appears to be one of the youngest protostars in Taurus.

\section{Description of Observations}\label{observations}

Observations of IRAM 04191 were obtained with the \emph{Spitzer Space Telescope} (Werner et al. 2004) as part of the Legacy Project ``From Molecular Cores to Planet Forming Disks'' (c2d; Evans et al. 2003).  Both the Infrared Array Camera (IRAC; Fazio et al. 2004) and the Multiband Imaging Photometer (MIPS; Rieke et al. 2004) were used.  The IRAC observations were obtained in all four bands, resulting in images at 3.6 (IRAC band 1), 4.5 (IRAC band 2), 5.8 (IRAC band 3), and 8.0 (IRAC band 4) \um.  Two epochs of observations were taken in order to enable the identification and removal of asteroids.  The first epoch was observed on 2004 September 8 (Program ID [PID] 139, AOR key 0005073920); the second was observed on 2004 September 10 (PID 139, AOR key 0005074432).  The observations in each epoch consisted of two 12 s images, dithered by $\sim 10 \as$, resulting in approximately a $5 \am \times 5 \am$ field.  They each also included a 0.6 s ``High-Dynamic Range'' (HDR) mode image.

The MIPS observations were obtained in the first two bands of the instrument (24 and 70 \um).  As with IRAC, two epochs of observations were taken to identify and remove asteroids.  The first epoch was observed on 2004 September 22 (PID 139, AOR key 0009416960); the second was observed on 2004 September 24 (PID 139, AOR key 0009411328).  The integration times were 36 and 100 s at 24 and 70 \um, respectively.  The sky offset position was $300\as$ in the scan direction.

The IRAC and MIPS images were processed by the Spitzer Science Center (SSC), using their standard pipeline, version S11, to produce Basic Calibrated Data (BCD) images.  These images were then improved by the c2d Legacy project to correct for artifacts.  A complete description of the improvements made can be found in the c2d data delivery documentation (Evans et al. 2006), as well as in Harvey et al. (2006) for the IRAC data and Young et al. (2004b) for the MIPS data.  After correcting for artifacts, mosaics were produced for the IRAC and MIPS images using the MOPEX software provided by the SSC.  Photometry was then obtained using a modified version of DoPHOT (Schechter et al. 1993) that utilizes a digitized rather than analytic point source profile to better match the real \emph{Spitzer} data, as well as incorporating several other changes (a complete description of the modifications is given in Harvey et al. 2006).

We also include 350 \um\ observations from Wu et al. (in preparation), as well as the observations presented by \andre\ et al. (1999), which include detections of IRAM 04191 at wavelengths ranging from 60 to 1300 \um\ using the ISOPHOT instrument on ISO, the JCMT, and the IRAM 30 m telescope, and upper limits ranging from 2.1 to 25 \um\ from ISOCAM, ISOPHOT, and the observations presented in Hodapp (1994). Also included are upper limits from the Two Micron All Sky Survey (2MASS) at 1.25, 1.65, and 2.17 \um.

\section{Results}\label{results}

\subsection{IRAM 04191+1522}\label{iramresults}

\begin{figure}[t]
\plotone{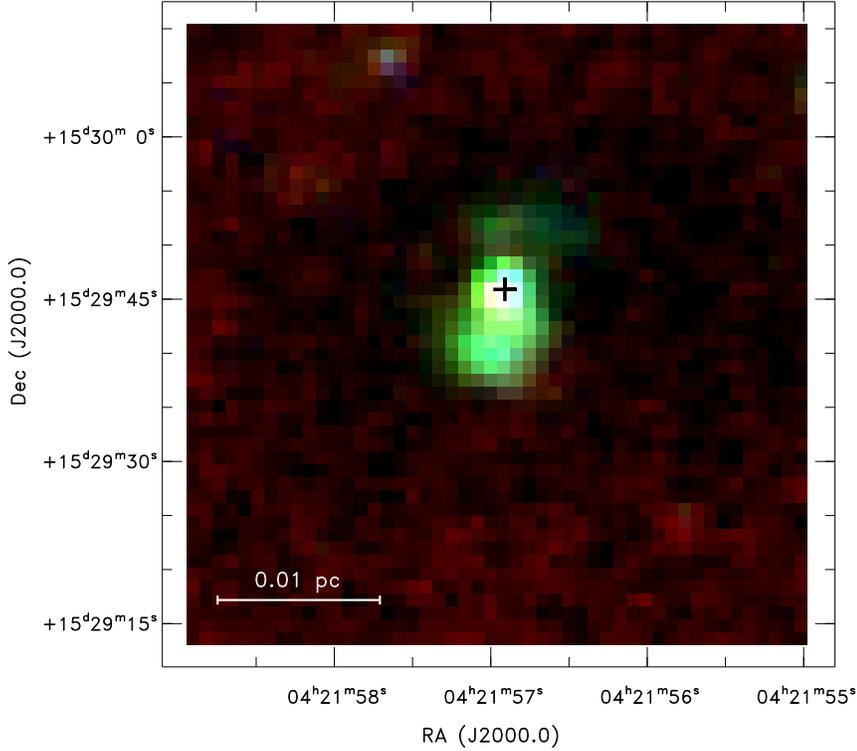}
\caption{\label{3color}Three-color image of IRAM 04191+1522 comprised of IRAC 1 (blue, 3.6 \um), IRAC 2 (green, 4.5 \um), and IRAC 4 (red, 8.0 \um).  The color scales are displayed using a linear stretch based on the pixel-to-pixel dispersion, $\sigma$, in the images ($\sigma_{3.6} = 0.9$ $\mu$Jy; $\sigma_{4.5} = 0.9$ $\mu$Jy; $\sigma_{8.0} = 4.1$ $\mu$Jy), with ranges [+3,+20]$\sigma$ for the 3.6 and 4.5 \um\ images and [-1,+7]$\sigma$ for the 8.0 \um\ image.  The black cross marks the position of the infrared point source IRAM 04191-IRS.  The green, extended emission seen in the vicinity of IRAM 04191-IRS likely arises from an outflow cavity, and the above ranges were chosen to emphasize this nebulosity while also showing the more diffuse, background emission present in band 4 (8.0 \um) due to PAH emission.}
\end{figure}

\begin{figure}[t]
\plotone{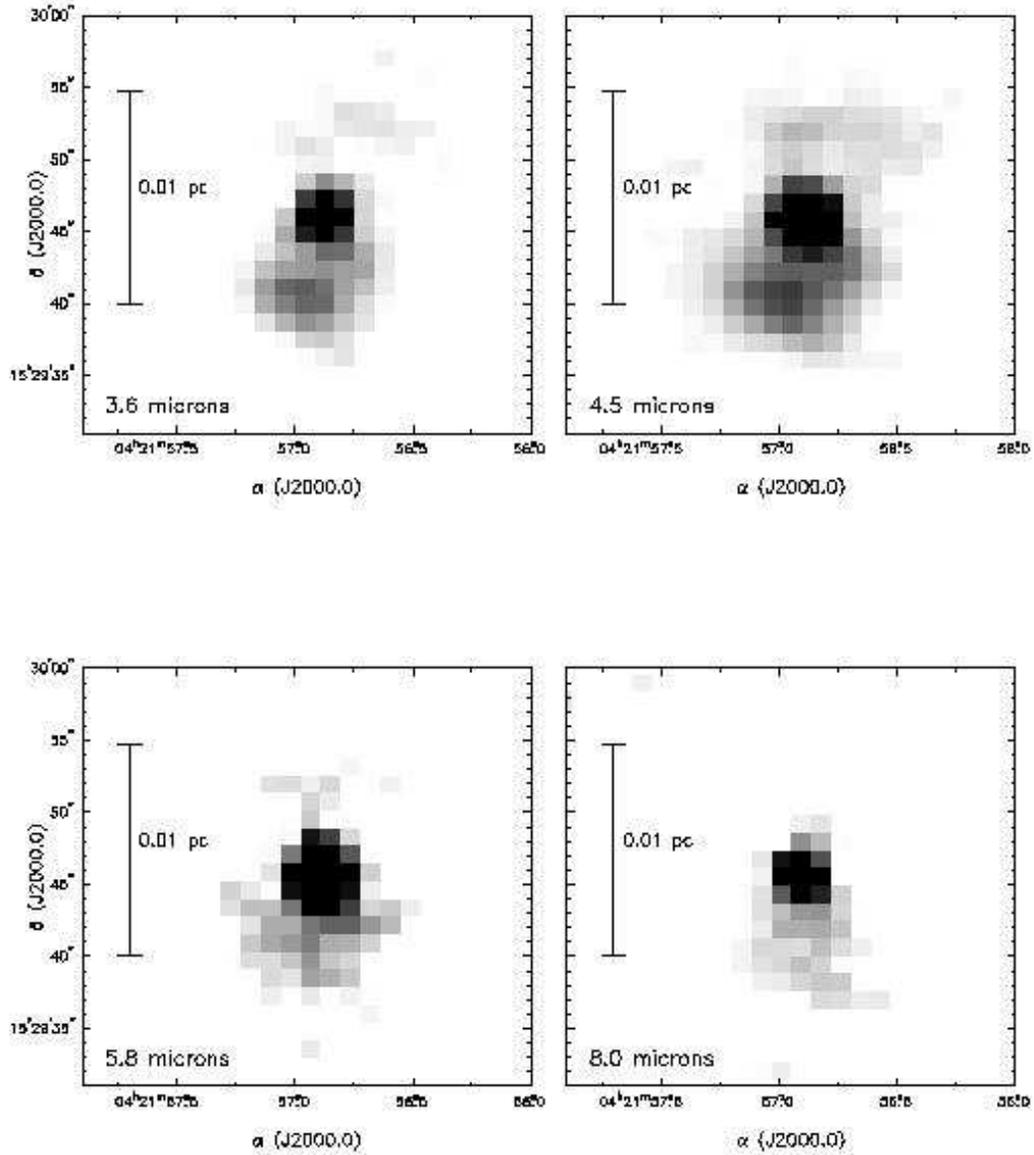}
\caption{\label{fourpanel}IRAC 3.6, 4.5, 5.8, and 8.0 \um\ images of IRAM 04191-IRS (left to right, top to bottom).  The images are displayed using a linear stretch based on the pixel-to-pixel dispersion, $\sigma$, in the images ($\sigma_{3.6} = 0.9$ $\mu$Jy; $\sigma_{4.5} = 0.9$ $\mu$Jy; $\sigma_{5.8} = 3.0$ $\mu$Jy; $\sigma_{8.0} = 4.1$ $\mu$Jy), with ranges [+4,+20]$\sigma$ for the 3.6 \um\ image, [+4,+30]$\sigma$ for the 4.5 \um\ image, [+3,+10]$\sigma$ for the 5.8 \um\ image, and [+4,+10]$\sigma$ for the 8.0 \um\ image.  These ranges were chosen to highlight the extended nebulosity associated with the point source.}
\end{figure}

An infrared point source, SSTc2d J042156.91+152945.9 (J2000 HHMMSS.ss+DDMMSS.s; hereafter referred to as IRAM 04191-IRS), is detected at all six wavelengths observed by \emph{Spitzer} (3.6, 4.5, 5.8, 8.0, 24, and 70 \um).  While the previously obtained 60 \um\ ISO detection (\andre\ et al. 1999) and the new 70 \um\ \emph{Spitzer} detection arise primarily from dust thermal emission (see \S \ref{lint}), the detections from $3.6 - 24$ \um\ represent the first direct detection of the embedded source itself.  Figure \ref{3color} shows a three-color image of IRAM 04191 comprised of IRAC band 1 (blue), IRAC band 2 (green), and IRAC band 4 (red).  Extended nebulosity is associated with the point source.  Figure \ref{fourpanel} displays all four IRAC images, and the extended nebulosity is seen at all four wavelengths.  It generally decreases in strength with increasing wavelength, as would be expected from scattered light, but it is stronger in band 2 (4.5 \um) than band 1 (3.6 \um) and thus appears green in Figure \ref{3color}.

\begin{figure}[t]
\plotone{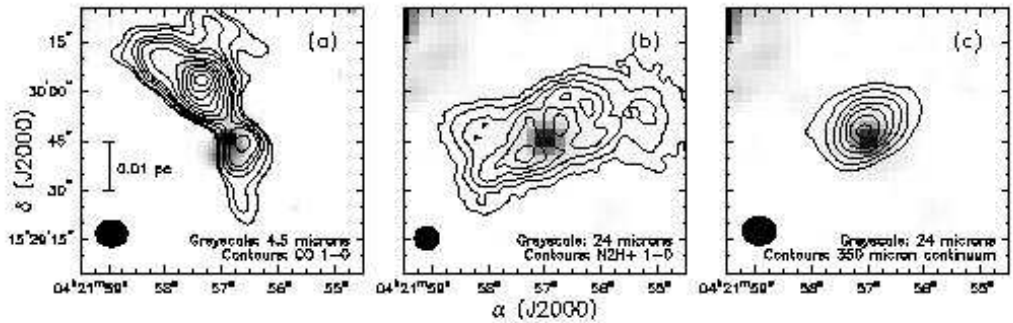}
\caption{\label{results2}\emph{Spitzer} images of IRAM 04191 overlaid with molecular line and continuum emission.  In each panel, the solid circle shows the beam for the emission shown in that panel.  (a) IRAC 2 (4.5 \um, greyscale) image of IRAM 04191 with contours showing the CO 1-0 emission from Lee et al. (2002), integrated over all velocities.  The contour levels are 15,20,25,30,35,45,...,95\% of the peak. (b) MIPS 1 (24 \um, greyscale) image of IRAM 04191, with contours showing the \nthp\ 1-0 emission from Lee et al. (2005), integrated over all velocities.  The contours start at 15\% of the peak and increase by 15\%.  (c) MIPS 1 (24 \um, greyscale) image of IRAM 04191 with contours showing the 350 \um\ continuum emission from Wu et al. (in preparation), starting at 15\% of the peak and increasing by 10\%.  The 350 \um\ continuum emission has a peak flux and noise level of 1.6 and 0.1 Jy beam$^{-1}$, respectively (Wu et al. in preparation).  }
\end{figure}

Figure \ref{results2}a shows the IRAC 2 (4.5 \um) image of IRAM 04191 overlaid with CO 1-0 contours from Lee et al. (2002) that trace the molecular outflow, integrated over all velocities.  The lobe to the south is blue-shifted while the lobe to the north is red-shifted.  The nebulosity associated with IRAM 04191-IRS is seen to correlate with the edge of the blueshifted outflow emission.  Combining this with the fact that the nebulosity is strongest in IRAC 2, a photometric band that includes H$_2$ emission lines excited by outflows, the nebulosity most likely originates from a combination of shocked emission from the molecular outflow and scattered light off the edge of an outflow cavity.  The redshifted CO emission is brighter than the blueshifted emission by about a factor of 2, similar to the outflow detected around L1014-IRS (Bourke et al. 2005).  Bourke et al. argued the observed brightness asymmetry in that outflow might arise because L1014-IRS is offset from the density peak, with the redshifted outflow propagating into denser material than the blueshifted emission and thus sweeping up a larger amount of mass.  A similar explanation may hold here, although the case for an offset between IRAM 04191-IRS and the density peak remains ambiguous (see below).

Figures \ref{results2}b and \ref{results2}c show the MIPS 1 (24 \um) image of IRAM 04191 overlaid with \nthp\ 1-0 contours from Lee et al. (2005), integrated over all velocities, and SHARC-II 350 \um\ continuum contours from Wu et al. (in preparation), both tracers of the circumstellar envelope.  The 350 \um\ continuum emission does not show as flattened a morphology as the \nthp\ 1-0 emission or the longer-wavelength continuum emission (e.g., Andre et al. 1999).  This is an expected result due to the presence of the outflow, as 350 \um\ continuum emission is more sensitive to the combination of temperature and column density than the other tracers, which are mostly sensitive only to the column density.  Thus the outflow, which is heating the dust perpendicular to the major axis of elongation, will remove some of the flattening in the emission at this wavelength.

The \emph{Spitzer} position of IRAM 04191-IRS given above is a weighted average of the position of the point source in each band; there is no systematic shift in the position of the source between bands.  This position agrees to within less than 0.2\as\ with the position given by Belloche et al. (2002) based on 227 GHz continuum emission detected by the IRAM Plateau de Bure Interferometer (PdBI).  Given that their reported position is accurate to within $\sim 0.5$\as, the positions of the infrared point source detected by \emph{Spitzer} and the millimeter continuum emission detected by the PdBI are consistent with each other.  However, as seen in Figure \ref{results2}, the center of the outflow and the center of the envelope as traced by the \nthp\ 1-0 emission do not exactly coincide with IRAM 04191-IRS itself but lie $\sim$ $4\as$ to the northwest, as first reported by Lee et al. (2005).  The 350 \um\ continuum emission also shows an offset to the north of $\sim 2-4\as$, but this is approximately the same size as the SHARC-II pointing uncertainty and thus not considered to be significant.

These observed offsets suggest that the embedded protostar detected by \emph{Spitzer} is offset from the center of the dense core in which it is embedded, yet the agreement between the \emph{Spitzer} and PdBI positions suggests otherwise.  This inconsistency could be resolved by assuming the 227 GHz continuum emission arises from a circumstellar disk rather than from the inner regions of the dense core, but Belloche et al. (2002) conclude that this is unlikely (see discussion in \S \ref{1d}).  Thus, it remains rather uncertain whether or not IRAM 04191-IRS is actually offset from the density peak of the surrounding material.  However, if the offset is real it could provide a possible explanation for the very low luminosity of this object.

Similar offsets are seen in other low luminosity sources (e.g., L1014, Young et al. 2004a), raising the possibility that these sources are perhaps drifting out of the regions of highest density before they can finish accreting from the surrounding envelope, as suggested by Huard et al. (2006).  An offset of $4\as$ at 140 pc corresponds to a distance of approximately 560 AU, and if we assume IRAM 04191-IRS has been moving relative to the core in the plane of the sky at a constant velocity for the dynamical time of the outflow ($0.8 \times 10^4$ years, \andre\ et al. 1999), this results in a relative velocity between the core and protostar of $\sim 0.35$ km s$^{-1}$.  For comparison, Huard et al. (2006) derived a projected relative velocity between L1014-IRS and the surrounding core of $\sim 0.1$ km s$^{-1}$, based on deep near-infrared extinction mapping of the core.

A drift velocity between IRAM 04191-IRS and its core of $\sim 0.35$ km s$^{-1}$ is consistent with the suggestion by Stamatellos et al. (2005) that asymmetries in the pattern of accretion onto a Class 0 protostar can give it a velocity of $\sim 0.3$ km s$^{-1}$ relative to its core.  Walsh, Myers, \& Burton (2004), on the other hand, concluded that such high velocities are not generally seen in low-mass star-forming cores.  Their conclusion was based in part on the fact that most protostellar sources were found to have very small offsets from the high density core.  The observed offset of $\sim 4\as$ would place IRAM 04191-IRS in the first bin of Figure 3 of Walsh, Myers, \& Burton (2004), which plots the distribution of fractional offsets of the protostars in their sample relative to the size of the core (defined to be the average size of the 50\% peak \nthp contour).  Since most of their sources fall within this bin, IRAM 04191-IRS has a fractional offset consistent with their findings.  The difference, however, is in the assumed age.  They assumed the protostars in their sample moved a typical distance of one core radius in 1 Myr, and from this derived typical drift velocities of 0.09 km s$^{-1}$.  

If the dynamical time of the outflow, used here to derive the drift velocity of IRAM 04191-IRS, is considered as a lower limit instead of the actual age of this source, the velocity may in fact be smaller and more consistent with the findings of Walsh, Myers, \& Burton (2004).  However, the very low luminosity of IRAM 04191-IRS potentially makes it a very different type of object than the normal, low-mass protostars studied by Walsh, Myers, \& Burton.  If the high drift velocity does imply that the source is moving out of the core before it can finish accreting the surrounding material, the study by Walsh, Myers, \& Burton may not have found such velocities simply because they did not include objects with such low luminosities as IRAM 04191-IRS.

Ultimately, the nature of this offset remains ambiguous.  The agreement between the \emph{Spitzer} and PdBI positions is hard to explain if IRAM 04191-IRS truly is moving out of its core at a relatively high velocity.  Lee et al. (2005) argue the observed offset could be explained by a previously unseen binary companion, but no evidence of such a companion is seen in the \emph{Spitzer} observations.  The situation is also complicated by the fact that the PdBI continuum emission is tracing the dust density peak while the molecular line peak is dependent not only on density but also on abundance (e.g., Lee, Bergin, \& Evans 2004), excitation, and source geometry.  Future, higher-resolution interferometer observations are required in order to further probe the inner, dense regions and discern the true position of the peak of the dense core relative to IRAM 04191-IRS.

Table \ref{iramphot} presents all existing photometry on IRAM 04191.  It lists the wavelength, flux density, uncertainty in flux density, aperture diameter, and reference for all photometric detections of this source, including both those from existing observations and the new \emph{Spitzer} observations.  The \emph{Spitzer} flux density uncertainties include both the statistical measurement uncertainties and an absolute calibration uncertainty of 15\% for the 4 IRAC bands and the MIPS 1 band ($3.6-24$ \um) (e.g., Harvey et al. 2006; Evans et al. 2006) and 20\% for the MIPS 2 band (70 \um).  Upper limits based on non-detections at various wavelengths are also listed.  Based on the observed SED from 3.6 to 1300 \um\ listed in this table, we calculate $\lobs = 0.13 \pm 0.03$ $\lsun$, $\tbol = 25 \pm 5$ K, and $\lobssmm = 10 \pm 5$ (where $\lsmm$ is defined to be the observed luminosity longward of 350 \um).  \lobs\ is the observed luminosity and is calculated by integrating the observed SED.  This quantity is usually referred to as the bolometric luminosity \lbol, as in \andre\ et al. (1999), and our calculated \lobs\ is in good agreement with their result.  We call this the observed luminosity instead of the bolometric luminosity because, for heavily embedded objects, the long-wavelength emission dominates the bolometric luminosity, and if the source is more extended at these wavelengths than the sizes of the apertures used for photometry, the observed luminosity will be less than the bolometric luminosity.

Because heating from the ISRF can provide up to a few tenths of a solar luminosity to the bolometric luminosity of an embedded source, for a low luminosity object such as IRAM 04191, the bolometric luminosity can be dominated by heating from the ISRF.  It is important to subtract this component from the total luminosity in order to get a true estimate of the internal luminosity.  This will be examined in detail in the following section using radiative transfer models, but we motivate this with a simple calculation.  Assuming that all of the long-wavelength ($\lambda > 100$ \um) emission arises from dust heated by the ISRF, an estimate of the internal luminosity can be obtained by integrating the observed SED for all wavelengths less than this, resulting in $\lint\ \sim 0.02$ \lsun.  This would qualify IRAM 04191-IRS as a VeLLO based on the definition given in \S \ref{intro}.  While the true value of the internal luminosity is likely to be higher since at least a fraction of the long-wavelength emission probably does arise from dust heated by the internal source, this calculation suggests that the internal luminosity for this object is indeed very low.

\subsection{IRAS 04191+1523}\label{irasresults}

As previously mentioned, the Class I source IRAS 04191 is located approximately 1\am\ to the northeast of IRAM 04191.  The position of this source was covered in both our IRAC and MIPS observations of IRAM 04191.  It is known to be a binary system from near-infrared studies (\duchene\ et al. 2004), and its two components are easily resolved by IRAC.  Source B, the fainter of the two components, is located to the northwest of Source A, and from the IRAC observations  we measure an angular separation of 6.5\as\ and a position angle for Source B of $304^{\circ}$, consistent with the values of 6.09\as\ and $303.7^{\circ}$ found for the angular separation and position angle, respectively, by \duchene\ et al. in the near-infrared.  The FWHM of the \emph{Spitzer} point-spread profile is  6.0\as\ at 24 \um\ and 18.0\as\ at 70 \um; thus the two sources, separated by 6.5\as, are just beyond the resolving limit at 24 \um\ and not at all resolved at 70 \um.  The emission at 24 \um\ does appear extended, suggesting the presence of a binary, but the two sources are fully resolved only at $\lambda \leq 8.0$ \um.  At a distance of 140 pc, an angular separation of 6.5\as\ corresponds to a minimum separation of approximately 910 AU, assuming no separation along the line of sight.  Table \ref{irasphot} lists the detections of these two components from both the new \emph{Spitzer} observations and from 2MASS, along with those of the single, unresolved system at 24 and 70 \um\ from the new \emph{Spitzer} observations, 450 and 850 \um\ from the SCUBA array on the JCMT (Young et al. 2006), and 1300 \um\ from the MAMBO array on the IRAM 30 m telescope (Motte \& \andre\ 2001).

\begin{figure}[t]
\plotone{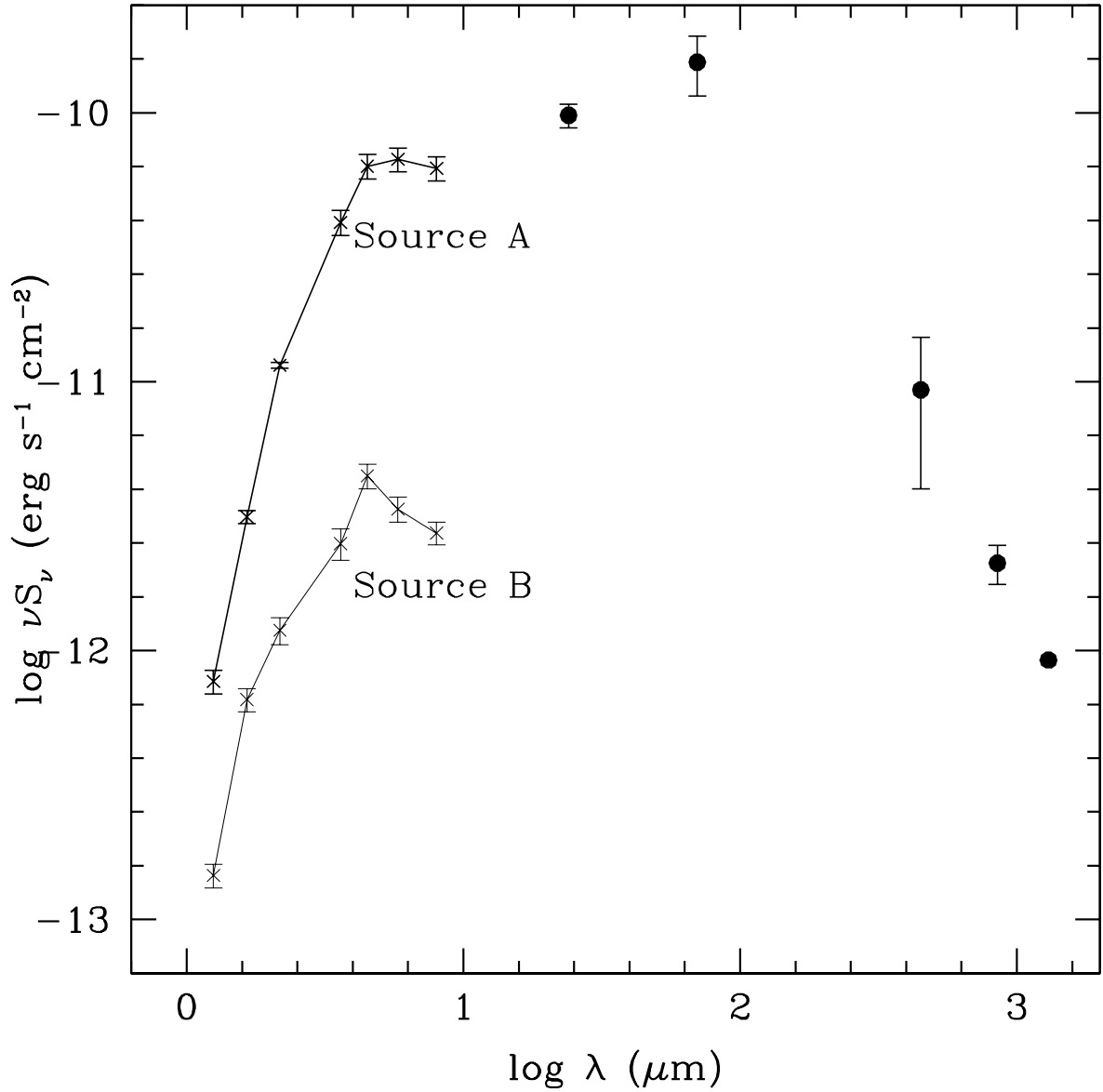}
\caption{\label{results3}SEDs of IRAS 04191+1523-A and IRAS 04191+1523-B, comprised of 2MASS and IRAC detections.  Sources A and B are labeled, and Source A is plotted with darker symbols.  Also shown are MIPS (24 and 70 \um), SCUBA (450 and 850 \um), and MAMBO (1300 \um) detections of the unresolved system.}
\end{figure}

Figure \ref{results3} shows the SEDs for both sources as well as for the unresolved system.  Source A appears to dominate the emission at the long wavelengths.  Associating all of the long-wavelength emission with Source A results in an observed luminosity for this source of $\lobs \sim 0.3$ $\lsun.$  However, the SCUBA fluxes are measured in 40\as\ diameter apertures and there are no observations between 70 and 450 \um, the wavelength range where the SED is expected to peak.  Thus, the bolometric luminosity is likely to be somewhat higher than the observed luminosity quoted above.  Based on the models of IRAM 04191 presented in \S \ref{models}, which take into account the sizes of the long-wavelength apertures, we estimate that the bolometric luminosity is likely to be at least a factor of $2-3$ higher than the observed luminosity.  For Source B, we calculate a luminosity including all detections at wavelengths $\leq 8.0$ \um\ of $L \sim 3 \times 10^{-3}$ \lsun.  The same calculation for IRAM 04191 yields $L \sim 2 \times 10^{-4}$ \lsun, demonstrating that Source B is more than an order of magnitude more luminous in the near- and mid-infrared than IRAM 04191.

\section{Radiative Transfer Models}\label{models}

In an effort to uncover the physical nature of IRAM 04191, in particular the internal luminosity of the embedded source, we have constructed physical models of this source, constrained by the SED.

\subsection{1-D Models}\label{1d}

We first modeled IRAM 04191 in one dimension in order to gain as much insight into the nature of the source as possible without the added complications of extra dimensions.  We used the 1-D radiative transfer package DUSTY (Ivezic et al. 1999) to calculate the temperature profile and observed SED of an internal source embedded in an envelope of material.  The program ObsSphere (Shirley et al. 2002) was then used to simulate the observational resolution at wavelengths longward of 100 \um\ where the source is more extended than the apertures used for photometry.  We assumed the dust opacities of Ossenkopf and Henning (1994) appropriate for thin ice mantles after $10^5$ years of coagulation at a gas density of $10^6$ cm$^{-3}$ (OH5 dust), which previous work has shown to be appropriate for cold, dense cores (e.g., Evans et al. 2001; Shirley et al. 2005).

The envelope of IRAM 04191 is heated by both an internal source and the ISRF.  For the ISRF we adopt that of Black (1994), modified at the ultraviolet wavelengths to reproduce the ISRF of Draine (1978).  This Black-Draine ISRF is then attenuated by $A_V = 3$ magnitudes of dust with properties given by Draine \& Lee (1984) to simulate being embedded in a parent cloud.  We assume a power-law radial density profile $[n(r) \propto r^{-p}]$ for the envelope with an index of $p=1.8$, which previous work has shown to provide a good fit to the observed 1.3 mm radial intensity profile (Motte \& \andre\ 2001).  The outer radius is set at 14,000 AU following the results of Motte \& \andre.  The inner radius is constrained to be less than about 260 AU, based on the conclusion by Belloche et al. (2002) that 227 GHz IRAM PdBI emission detected within the central 1.9\as\ ($\sim 260$ AU at 140 pc) is consistent with arising from the envelope rather than from a disk.  In our models we treat this inner radius, which determines the optical depth given an opacity law and density profile, as a free parameter subject to this constraint.  The mass of the envelope is determined by the long-wavelength, optically thin emission.  Assuming a fixed opacity law and strength of the ISRF, the envelope mass is the only parameter that controls the fit to the data at 850 \um\ and 1.3 mm, and we find that the total mass of the envelope is $M_{env} \sim 2.5$ M$_\odot$, which agrees with the $M_{env} \sim 1.5$ M$_\odot$ result from Belloche et al. (2002) to within a factor of 2.  These models predict a mass within a 60\as\ diameter aperture of $\sim 0.8$ \msun, in good agreement with \andre\ et al. (1999), who derived a mass of $\sim 0.5$ \msun\ within a radius of 4200 AU (corresponding to a radius of 30\as, or a diameter of 60\as, at $d=140$ pc).

While external heating of the envelope by the ISRF can explain most of the emission longward of 100 \um, the mid-infrared \emph{Spitzer} data requires the presence of a warmer internal source.  The simplest model is one in which a blackbody is used as the spectrum of the internal source (the input spectrum to DUSTY) in order to simulate a star.  While this can reproduce the flux observed from $3.6-8.0$ \um, it underestimates the flux detected at 24 \um\ by at least an order of magnitude, suggesting the presence of a cooler component to the input spectrum, such as a circumstellar disk.  This suggestion is reinforced by the presence of the molecular outflow.

We include a disk in our input spectrum by adding the emission from a disk to that of a stellar blackbody.  To simulate the disk emission we follow the method of Butner et al. (1994), which uses a simple model that calculates the emission from a disk at a given inclination with a temperature profile $T(r) \propto r^{-q}$, where $q$ is chosen to be 0.5 to simulate a flared disk.  The emission from the disk, averaged over all inclinations, is then added to the emission from the star to form the final input spectrum (Butner et al. 1994).  Such a disk emits through the reprocessing of stellar radiation, but this model still underestimates the 24 \um\ observation by at least a factor of 3.  Thus, we also include a component to the emission from an intrinsic disk luminosity, perhaps generated by accretion within or onto the disk.

Averaging the disk emission over all inclinations is equivalent to a disk at an inclination of $i=60\mbox{$^{\circ}$}$ (Butner et al. 1994).  Whitney et al. (2003) showed that the 24 \um\ flux is particularly sensitive to the inclination, and while most of this variation occurs when an outflow cavity intersects with the line of sight and is thus not relevant to these 1-D models, variation is observed even at inclinations where the outflow cavity is not a factor.  The variation in these cases is due instead to the effects of observing the disk at different inclinations.  Thus, in theory, lower disk inclinations could resolve the discrepancy between the observed and modeled flux at 24 \um\ without requiring an intrinsic disk luminosity.  To test this, we calculated the emission from the disk at various inclinations and compared it to the disk emission after averaging over all inclinations.  The difference between the 24 \um\ disk emission averaged over all angles and the emission at an inclination of $i=20\mbox{$^{\circ}$}$ is only a factor of 1.2, and the total difference between the 24 \um\ disk emission at $i=70\mbox{$^{\circ}$}$ and at $i=20\mbox{$^{\circ}$}$ is only a factor of 1.8.  More extreme inclinations are highly improbable based on the bipolar nature of the outflow (\andre\ et al. 1999), and regardless, the difference between the emission at $i=20\mbox{$^{\circ}$}$ and $i=0\mbox{$^{\circ}$}$ is negligible.  Thus, lower disk inclinations alone are not sufficient to match the 24 \um\ observation; an intrinsic disk luminosity must be included.  Exactly how much of the resolution of the discrepancy between the observed and modeled 24 \um\ flux arises from an intrinsic disk luminosity and how much arises from a lower disk inclination is not constrained by these simple 1-D models, but since these models are concerned primarily with constraining the internal luminosity rather than the stellar and disk luminosities individually, this is not a cause for concern. 

The inner radius of the disk is set to be the radius at which the temperature is equal to the dust destruction temperature (assumed to be 2000 K).  The outer radius of the disk, $R_d$, is set to the centrifugal radius, the radius where infalling matter in the equatorial plane encounters a centrifugal barrier due to conservation of angular momentum.  Given the sound speed, $c_s$, the age of the object, $t$, and the angular velocity of the cloud prior to collapse, $\Omega_0$, the centrifugal radius is calculated as $R_C = \frac{m_0^3}{16}c_s t^3 \Omega_0^2$, where $m_0$ is a dimensionless constant of order unity (Terebey et al. 1984; Young \& Evans 2005).  Assuming that the sound speed includes both thermal and turbulent motions, and assuming $T=10$ K, $t=2 \times 10^4$ years (\andre\ et al. 1999), $\Omega_0=4 \times 10^{-13}$ s$^{-1}$ (\andre\ et al. 1999), and a turbulent velocity of $v_{turb}=0.085$ km s$^{-1}$ (Belloche et al. 2002), we calculate $R_C = 3.3$ AU.

A disk size equal to this centrifugal radius is consistent with the constraint by Belloche et al. (2002) that the disk size be less than 10 AU, which they derive by assuming all of the flux detected from the previously mentioned 227 GHz IRAM PdBI data originates from an optically thick disk.  However, this disk size is significantly smaller than what Harvey et al. (2003) found for B335, a Class 0 source with a similar centrifugal radius.  They found a disk size of $\sim$ 45 AU despite B335 having a centrifugal radius of 3 AU, and this size of 45 AU is similar to the sizes measured for Perseus Class 0 disks (Brown et al. 2000).  Although these sources are of higher luminosity than IRAM 04191 (e.g., $L \sim 3$ $\lsun$ for B335; Harvey et al. 2003), this might suggest the disk size assumed in this work is too small if not for the constraint by Belloche et al. that it be less than 10 AU in size.  Since this constraint was derived by assuming that \emph{all} of the flux detected in the central beam of their 227 GHz PdBI data originated from a disk, despite their claim that, in reality, the flux appears to originate from the envelope, it is a strong upper limit on the size of the disk.  This model disk has a mass of $M_d = 5 \times 10^{-3}$($R_d$/3.3 AU)$^{0.5}$ \msun, and predicts a flux of 1.6 mJy at 227 GHz.  Belloche et al. measure a peak 227 GHz PdBI flux of 6.1 mJy/1.9\as\ beam, thus the flux predicted from this disk is only $\sim$ 25\% of the total flux and is consistent with most of the flux originating from the envelope.

A grid of models was constructed to test the four free parameters: the temperature of the star ($T_s$), the luminosity of the star ($L_s$), the intrinsic luminosity of the disk ($L_d$), and the envelope inner radius ($r_i$).  In order to better compare the models and the observations, we used our models to simulate the observations.  At the short wavelengths this is necessary because the IRAC and MIPS photometric filters extend over features in the spectrum (such as the 10 \um\ silicate feature) that can have a strong effect on the flux detected, and at the long wavelengths it is necessary because the flux detected will depend on the aperture size used.  Thus, for each model the observations were simulated, at the short wavelengths ($3.6$ $\mu m \leq \lambda \leq 70$ $\mu m$) by convolving the total flux density from the models with the IRAC and MIPS photometric filters, and at the long wavelengths ($\lambda > 100$ $\mu m$) by using ObsSphere to integrate over the extended emission.  Two reduced $\chi^2$ values were then calculated:

\begin{equation}\label{chi1}
\chi^2_1 = \frac{1}{k}\displaystyle\sum_{i=0}^n \frac{[S_{\nu}^{obs}(\lambda_i)-S_{\nu}^{mod}(\lambda_i)]^2}{\sigma_{\nu}(\lambda_i)} \quad , \lambda_i = \emph{Spitzer}
\end{equation}
\begin{equation}\label{chi2}
\chi^2_2 = \frac{1}{k}\displaystyle\sum_{i=0}^n \frac{[S_{\nu}^{obs}(\lambda_i)-S_{\nu}^{mod}(\lambda_i)]^2}{\sigma_{\nu}(\lambda_i)} \quad , \lambda_i > 100 \mu m
\end{equation}

In these equations, $S_{\nu}^{obs}(\lambda_i)$ is the observed flux density at $\lambda_i$, $S_{\nu}^{mod}(\lambda_i)$ is the simulated observed flux density from the model at this wavelength, and $\sigma_{\nu}(\lambda_i)$ is the uncertainty in the observed flux density.  For $n$ data points and $m$ free parameters there are $k=n-m$ degrees of freedom in these models, and we divide by $k$ to obtain the reduced $\chi^2$ values.  $\chi^2_1$ is calculated for the \emph{Spitzer} observations ($n=6$), and $\chi^2_2$ is calculated for the observations longward of 100 \um\ ($n=5$)\footnote{The photometry at 350 \um\ was performed in a 40\as\ aperture while all of the other long-wavelength photometry was performed in 60\as\ apertures.  Thus, the 350 \um\ SHARC-II observations are not included in the calculation of $\chi^2_2$ in order to ensure a uniform sample.}.  The division between the two $\chi^2$ values is based both on the fact that it represents the division between new \emph{Spitzer} observations and existing observations from other facilities, and because it represents the division between emission dominated by internal heating and emission dominated by external heating (see \S \ref{lint}).

Regardless of the assumptions used for the internal source, the ISRF provides $\sim 0.2$ $\lsun$ to the bolometric luminosity of the source.  Thus the bolometric luminosity is not simply the luminosity of the star and disk, but instead $L_{bol} = L_s + L_d +0.2$.  For all of the models in the grid, $\chi^2_2$ is less than about 15, and, for most models, it is less than 10.  Thus, compared to the short wavelengths (see below), all of the models fit the long-wavelength emission reasonably well.  The long-wavelength emission is constrained primarily near the peak of the SED ($\lambda \sim 200$ \um) by the strength and attenuation of the ISRF, and at the longest wavelengths (850 and 1300 \um) by the envelope mass and dust opacities.  Thus, the fact that all of these models fit the long-wavelength emission at least reasonably well suggests the assumptions used for these parameters are reasonable, but variations should be explored in future work.

\begin{figure}[t]
\plotone{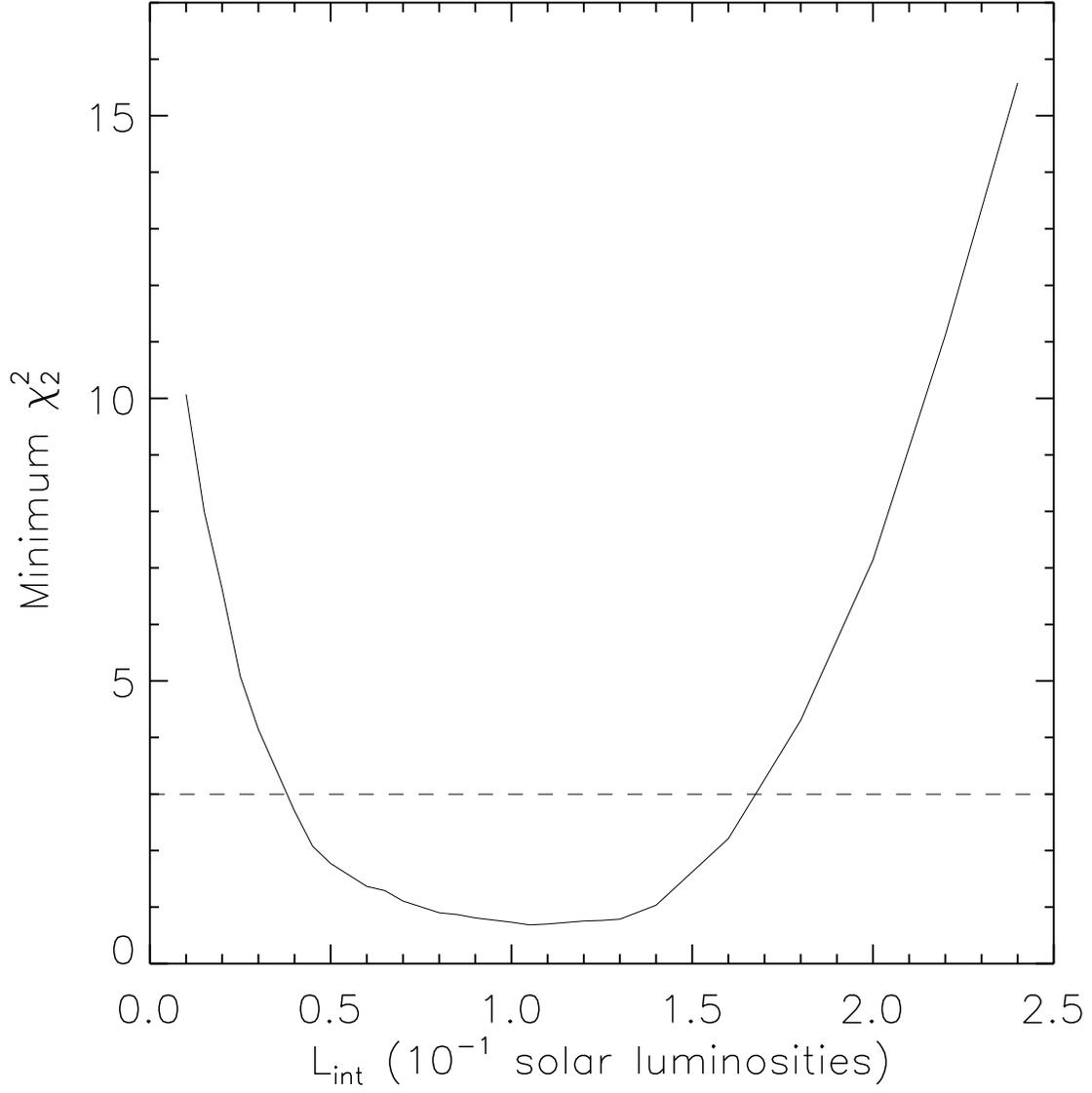}
\caption{\label{1dminchi2}Minimum value of $\chi^2_2$ for each value of \lint\ in the grid of 1-D models.  The dotted line represents the 95\% confidence limit.}
\end{figure}

Despite the fact that the emission longward of 100 \um\ is dominated by emission from external heating, the emission at 160 and 200 \um\ does still contain a small component from dust heated by the internal source.  This component is enough to set constraints on the internal luminosity of the central source, which, as before, is defined to be the luminosity of the central object (star+disk), $L_{int} = L_s + L_d$.  If the internal luminosity is too low the dust is not heated enough and the models predict too little emission at 160 and 200 \um\ compared to the data.  In the same manner, an internal luminosity that is too high results in too much emission at these wavelengths.  Figure \ref{1dminchi2} plots the minimum value of $\chi^2_2$ for each value of \lint\ in the grid of models.  Imposing a requirement of $\chi^2_2 \leq 3.0$, equivalent to a confidence level of approximately 95\%, constrains the internal luminosity to be in the range $0.04 \leq \lint \leq 0.16$ \lsun.  However, since this wavelength range is very sensitive to the strength and attenuation of the ISRF (e.g., Shirley et al. 2005), parameters that were held fixed in these models, these limits on \lint\ must be confirmed at the shorter wavelengths that are unaffected by the details of the ISRF.

$\chi^2_1$, a measure of the quality of the fit at the shorter wavelengths observed by \emph{Spitzer}, shows much more variation over the parameter space than $\chi^2_2$.  This demonstrates that the emission at the short wavelengths is much more sensitive to the details of the central source and central regions of the envelope than the emission at the longer wavelengths.  Despite showing variation over the grid of models, $\chi^2_1$ does not, however, show variation over $T_s$.  Thus, we assume a stellar temperature of 3000 K but note that our models do not constrain this parameter.

\begin{figure}[t]
\plotone{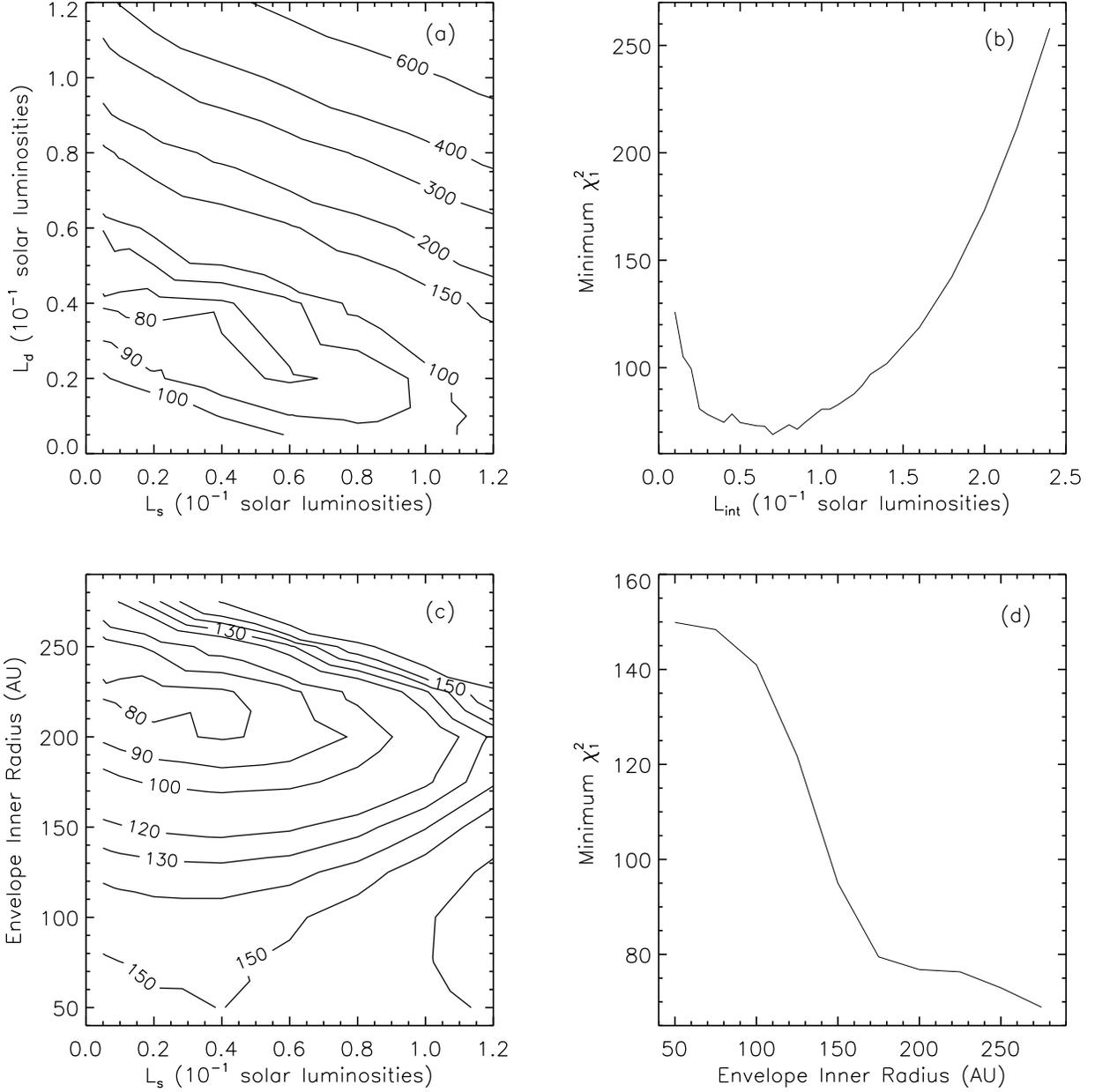}
\caption{\label{1dcontours}Results from the grid of 1-D models:  (a) $\chi^2_1$ contours over the grid of $L_s$ and $L_d$, with $T_s$ and $r_i$ held constant at 3000 K and 225 AU, respectively.  (b) Minimum value of $\chi^2_1$ for each value of \lint\ in the grid of models.  (c) $\chi^2_1$ contours over the grid of $L_s$ and $r_i$, with $T_s$ and $L_d$  held constant at 3000 K and 0.04 \lsun, respectively.  (d) Minimum value of $\chi^2_1$ for each value of $r_i$ in the grid of models.}
\end{figure}

Figure \ref{1dcontours}a shows $\chi^2_1$ contours over $L_s$ and $L_d$ assuming a fixed stellar temperature and envelope inner radius of 3000 K and 225 AU, respectively.  The slope of the contours show that the models constrain $L_s + L_d$ more than $L_s$ and $L_d$ individually, and since it is the internal luminosity that is the focus of these efforts, we do not attempt to separate the stellar and disk contributions to the internal luminosity.  From this figure, we constrain $L_{int} = 0.08 \pm 0.04$ $\lsun$.  In order to ensure that a ``cross section'' of models with the other parameters held constant, such as presented in Figure \ref{1dcontours}a, does not bias the results in any way, Figure \ref{1dcontours}b shows the minimum value of $\chi^2_1$ for each value of \lint\ in the grid of models, with no constraints on the other parameters.  This figure is consistent with $L_{int} = 0.08 \pm 0.04$ $\lsun$, a result that is also consistent with the constraint derived above from the fits to the long-wavelength emission.  An internal luminosity of $0.08 \pm 0.04$ $\lsun$ gives a bolometric luminosity of $L_{bol} \sim 0.28$ $\lsun$, about a factor of two higher than previously reported (\andre\ et al. 1999).  This difference arises because of the distinction between observed luminosity and bolometric luminosity as discussed in \S \ref{results}.  The value for \lbol\ presented here takes into account the fact that the source is more extended at the long wavelengths than the apertures used for photometry.

Figure \ref{1dcontours}c shows $\chi^2_1$ contours over $L_s$ and $r_i$ assuming a fixed stellar temperature and intrinsic disk luminosity of 3000 K and 0.04 $\lsun$, respectively.  The best models are found from this figure to be those with $r_i = 225 \pm 25$ AU.  However, to again ensure the results are not biased by holding the other parameters fixed, Figure \ref{1dcontours}d shows the minimum value of $\chi^2_1$ for each value of the envelope inner radius in the grid of models.  This figure demonstrates that the envelope inner radius is not actually as strongly constrained as suggested by Figure \ref{1dcontours}c.  A minimum envelope inner radius of $\sim 150$ AU is required in order to allow enough mid-infrared emission to escape to match the observations, but larger inner radii are also allowed because the resulting decrease in optical depth through the envelope can be offset by a lower internal luminosity, resulting in approximately the same amount of mid-infrared emission.  These models set a lower limit to the envelope inner radius of 150 AU, and combining this with the upper limit of 260 AU by Belloche et al. (2002) discussed above, we constrain $150 \leq r_i \leq 260$ AU.

\begin{figure}[t]
\plotone{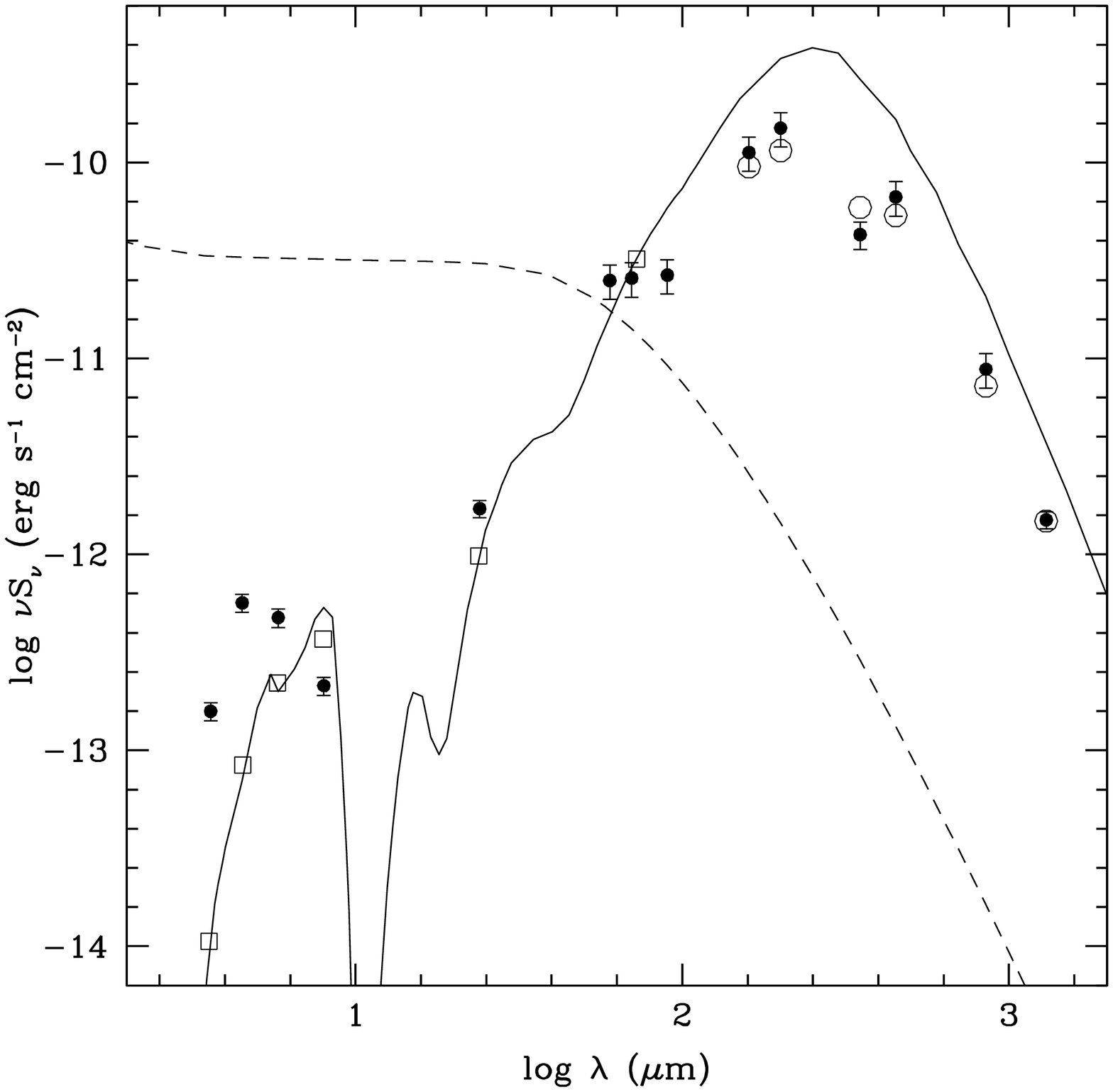}
\caption{\label{1dsed}Best-fit 1-D model for IRAM 04191+1522.  The solid line shows the monochromatic flux density predicted by the model while the dashed line shows the input spectrum.  The filled circles with error bars show the observed fluxes, the open squares show the results of convolving the model flux with the IRAC and MIPS photometric filters, and the open circles show the results of simulating the apertures used for photometry with ObsSphere.  A perfect-fit model would be one in which the open squares and circles are aligned with the filled circles.  The model parameters are $T_s = 3000$ K, $L_s = 0.04$ $\lsun$, $L_d = 0.04$ $\lsun$, and $r_i = 225$ AU (see text for discussion).}
\end{figure}

Figure \ref{1dsed} shows a representative model from the constraints given above.  The specific parameters used in this model are $T_s=3000$ K, $L_s=0.04$ $\lsun$, $L_d=0.04$ $\lsun$, and $r_i=225$ AU.  This model, despite being representative of the best fits obtained, is not a very ``good'' fit.  In fact, $\chi^2_1 = 79$ for the model presented in the figure.  The main problem is an inability to match the shape of the emission in the $3.6-8.0$ \um\ range.  Removing the disk could improve the fit at these wavelengths by changing the shape of the input spectrum, but, as already discussed, a disk is required to match the 24 \um\ data.

Clearly, the 1-D models presented above do not provide good fits to the observations at $3.6-8.0$ \um.  One of the more uncertain parameters in these models is the density radial profile index $p$, which we held fixed at $p = 1.8$ based on values reported in the literature (Motte \& \andre\ 2001).  If a circumstellar disk is present, it could lead to an artificially steep density profile by adding an unresolved component to the millimeter flux used to derive this value.  To address this we examined models with $p = 1.5$ but found they did not substantially improve the fits between the models and the data because, in the $3.6-8.0$ \um\ range, a change in $p$ will be balanced by a change in the inner radius of the envelope in order to keep the optical depth at these wavelengths approximately the same.  Thus, the fits at these wavelengths can not be significantly improved by allowing the radial density profile to vary.  Furthermore, since, in general, the SED is not sensitive to the detailed shape of the density profile (e.g., Evans et al. 2001; Shirley et al. 2002), our models do not constrain $p$ and we are justified in holding it fixed at $p = 1.8$.

\subsection{2-D Models}\label{2d}

In \S \ref{results} it was noted that there is strong evidence for the existence of an outflow cavity, a structure that can not be incorporated into a 1-D model.  This evidence, combined with the failure of our 1-D models to provide good fits to the data at $\lambda \leq 8.0$ \um, leads us to investigate 2-D models with outflow cavities.  To include these cavities in our models we used the two-dimensional Monte Carlo dust radiative transfer code RADMC (Dullemond \& Turolla 2000; Dullemond \& Dominik 2004).  In these models a conical outflow cavity of opening half-angle $\theta$ is cut out of the envelope and the density within the cavity is set to zero.  In all other respects, including the radial density profile, the dust opacities, the ISRF, and the envelope outer radius, the models are identical to the 1-D models discussed above.

Modeling in 2-D with conical outflow cavities introduces two new free parameters: the opening half-angle of the outflow cavity, $\theta$, and the inclination of the source, $i$.  As mentioned above, \andre\ et al. (1999) found the inclination to be $i \sim 50\mbox{$^{\circ}$}$.  Furthermore, the mid-infrared emission strongly depends on the inclination angle when an outflow cavity is present (e.g., Whitney et al. 2003), with the emission rising dramatically with decreasing inclination once the line of sight intersects the outflow cavity.  This leads us to set the opening half-angle of the outflow cavity to $50\mbox{$^{\circ}$}$ so that the outflow cavity will begin to affect the observed SED at $i \sim 50\mbox{$^{\circ}$}$.  This is also consistent with our attempts to measure the opening angle based on the extended emission seen in the IRAC images, from which we measure $\theta \sim$ $40\mbox{$^{\circ}$}-60\mbox{$^{\circ}$}$.  Thus, in all the 2-D modeling presented below we assume a conical outflow cavity with an opening half-angle of $50\mbox{$^{\circ}$}$ and allow the inclination to vary.  The effects of different opening half-angles and outflow cavity geometries are not explored here but should be considered in future work.

As with the 1-D models, we must decide how to treat the input spectrum.  Including only a star and no disk is ruled out due to the presence of the molecular outflow, and so we again include a disk.  We again find that the 24 \um\ observation is only matched if the disk emission includes a component from both reprocessing of stellar radiation and from an intrinsic luminosity.  Thus, for the input spectrum we used the emission from a star combined with the emission from a disk with an intrinsic luminosity, calculated following Butner et al. (1994) as described above.  A grid of models was then constructed to test the five free parameters:  the temperature of the star ($T_s$), the luminosity of the star ($L_s$), the intrinsic luminosity of the disk ($L_d$), the envelope inner radius ($r_i$), and the source inclination ($i$).

\begin{figure}[t]
\plotone{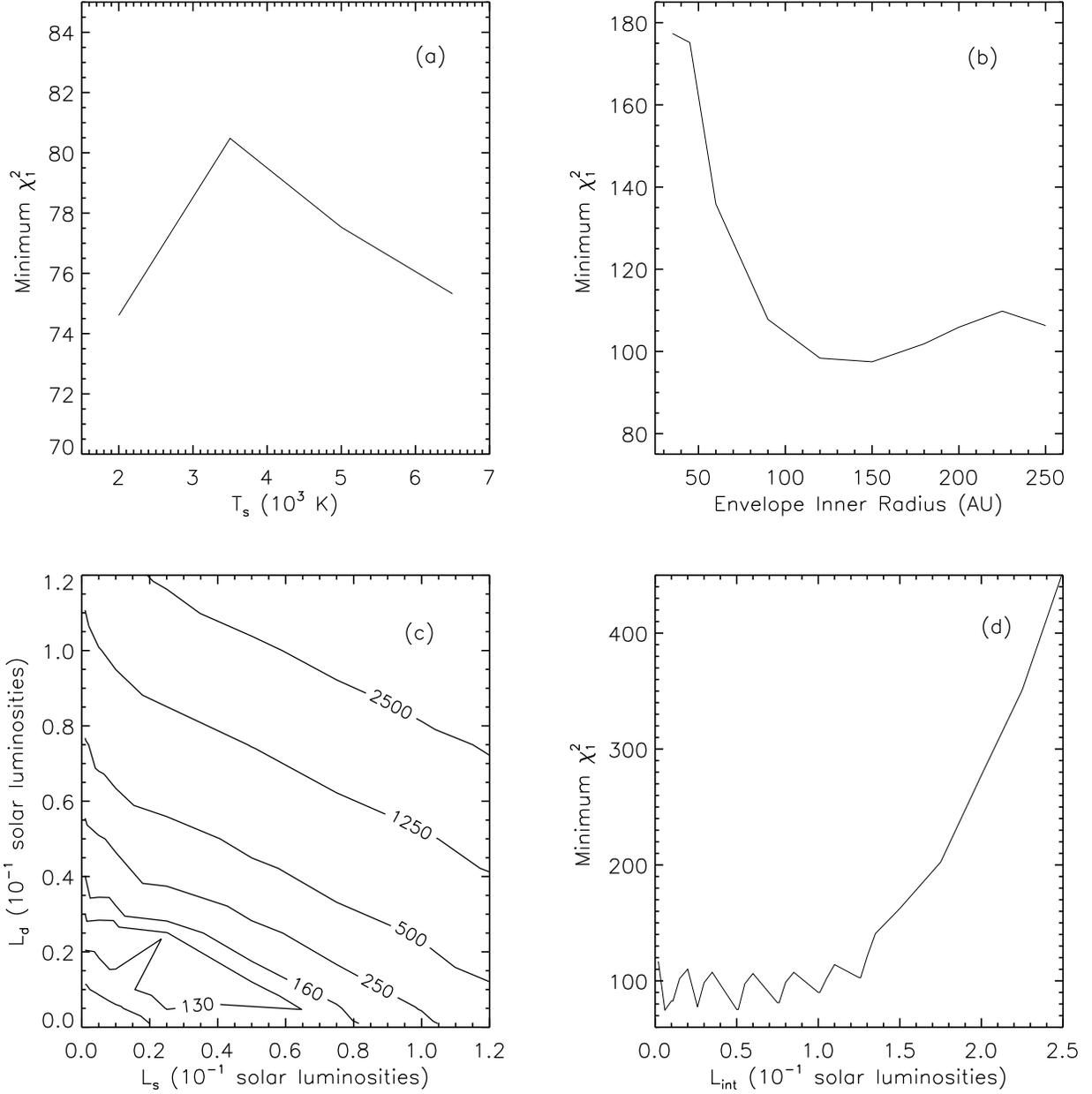}
\caption{\label{2dcontours}Results from the grid of 2-D models:  (a) Minimum value of $\chi^2_1$ for each value of $T_s$ in the grid of models.  (b) Minimum value of $\chi^2_1$ for each value of $r_i$ in the grid of models.  (c) $\chi^2_1$ contours over $L_s$ and $L_d$, with $T_s$, $r_i$, and $i$ held constant at 2000 K, 90 AU, and 49.5\mbox{$^{\circ}$}, respectively.  The spike seen in the $\chi^2_1=130$ contour at $L_s \sim L_d \sim 0.2$ $\lsun$ is a result of the steps used in the model grid and should be ignored.  (d) Minimum value of $\chi^2_1$ for each value of \lint\ in the grid of models.}
\end{figure}

Figure \ref{2dcontours}a shows the minimum value of $\chi^2_1$ for each value of $T_s$ in the grid of 2-D models.  The models are not very sensitive to the stellar temperature and thus we do not constrain this parameter.  However, unlike the 1-D models which showed no variation with $T_s$, the 2-D models are somewhat influenced by this parameter (see below).  To determine the constraints on $r_i$, we use the result from \S \ref{1d} that displaying $\chi^2_1$ contours over $r_i$ with the other parameters held fixed can bias the results, and instead plot in Figure \ref{2dcontours}b the minimum value of $\chi^2_1$ for each value of the envelope inner radius in the grid.  A minimum envelope inner radius of $\sim 90$ AU is required.  This is smaller than the minimum envelope inner radius of 150 AU required by the 1-D models because the outflow cavities allow more mid-infrared emission to escape for the same inner radius.  As before, larger inner radii are allowed because the decreased optical depth is offset by a decrease in internal luminosity.  Larger envelope inner radii are also allowed in these 2-D models because an object with a larger inner radius but also a larger inclination angle will produce about the same amount of mid-infrared emission.  This is evident in these models: the best models for $r_i \leq 200$ all had an inclination $i =$ 49.5\mbox{$^{\circ}$}, while the best models for $r_i > 200$ AU all had an inclination $i =$ 50\mbox{$^{\circ}$}.  Combining the above lower limit to the envelope inner radius with the upper limit of 260 AU by Belloche et al. (2002) discussed in \S \ref{1d}, we constrain $90 \leq r_i \leq 260$ AU.

Figure \ref{2dcontours}c shows contours over $L_s$ and $L_d$, with $T_s$, $r_i$, and $i$ held constant at 2000 K, 90 AU, and 49.5\mbox{$^{\circ}$}, respectively.  Similar to the 1-D models, the slope of these contours shows that the models constrain the internal luminosity more than the stellar and disk luminosities individually, and we constrain $L_{int} = 0.05 \pm 0.03$ $\lsun$.  To ensure that holding the other parameters fixed in this contour plot does not bias the results, Figure \ref{2dcontours}d plots the minimum value of $\chi^2_1$ for each value of \lint\ in the grid.  From this figure we constrain $\lint \leq$ 0.12 $\lsun$.

Higher internal luminosities than suggested by the contours in Figure \ref{2dcontours}c are allowed because $T_s$ is allowed to vary instead of being held fixed at 2000 K.  Because the outflow cavities allow more mid-infrared emission to escape directly through the envelope than the spherical, 1-D models, the actual shape of the input SED (and thus the value of $T_s$) does have an impact on the results.  Specifically, higher values of $T_s$ shift the peak of the input SED towards shorter wavelengths, causing a greater fraction of the stellar radiation to be reprocessed to longer wavelengths.  Thus, a larger value of \lint\ is required to match the observed amount of mid-infrared emission.  Since these 2-D models do not actually constrain the stellar temperature, these higher values of \lint\ can not be ruled out.  Figure \ref{2dcontours}d also suggests that very small internal luminosities are allowed, but, in reality, \lint\ must be greater than $\sim 0.04$ \lsun\ (see discussion below).

\begin{figure}[t]
\plotone{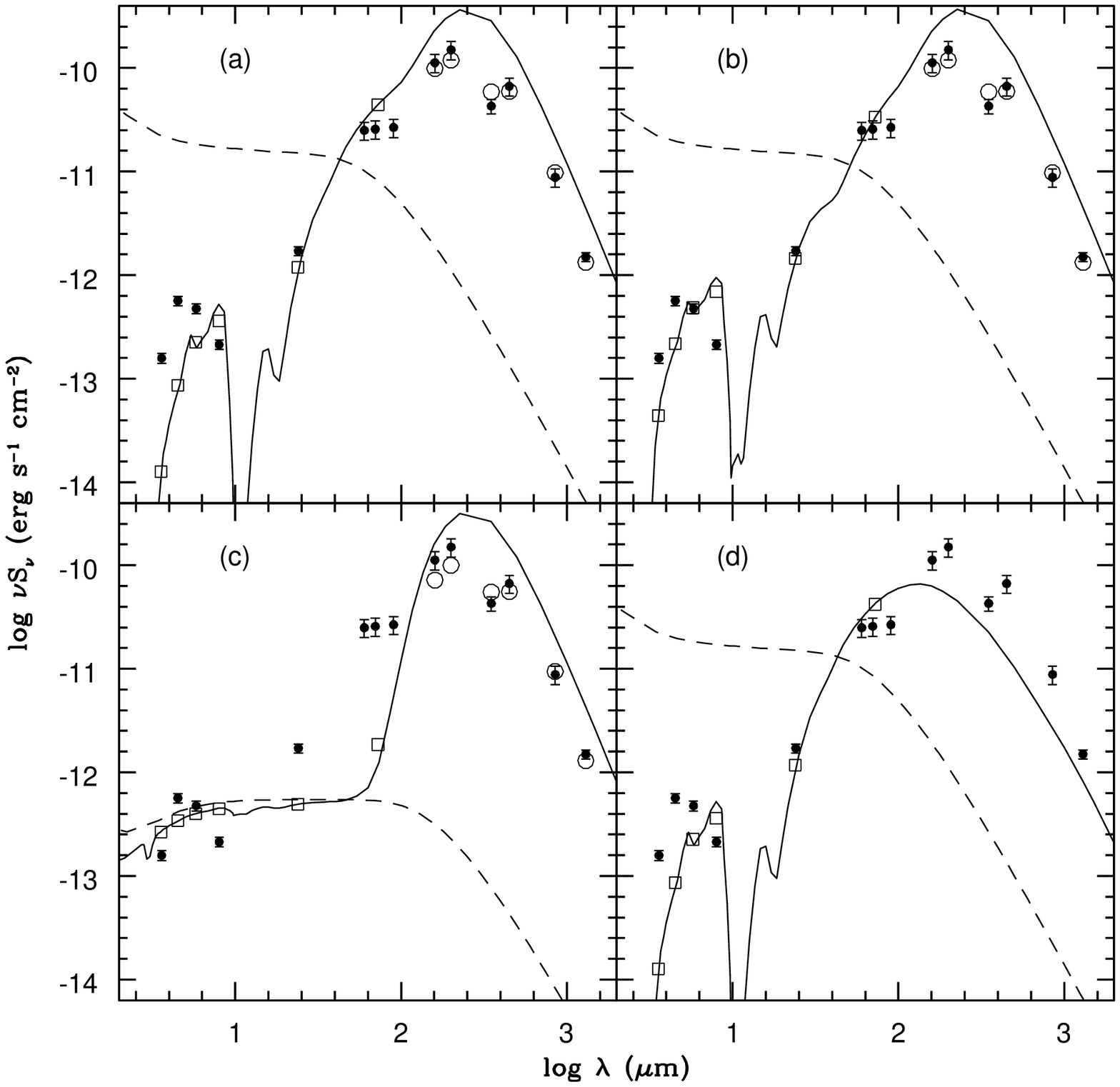}
\caption{\label{2dsed}2-D models of IRAM 04191+1522.  In each panel the solid line shows the monochromatic flux density predicted by the model while the dashed line shows the input spectrum.  The filled circles with error bars show the observed fluxes, the open squares show the results of convolving the model flux with the IRAC and MIPS photometric filters, and the open circles show the results of simulating the apertures used for photometry.  (a) A best-fit model: $T_s = 3500$ K, $L_s = 0.06$ $\lsun$, $L_d = 0.02$ $\lsun$, $r_i = 90$ AU, and $i = 49.5\mbox{$^{\circ}$}$.  (b) Same as (a), except with $r_i = 260$ AU and $i = 50.0\mbox{$^{\circ}$}$.  (c) The best model for $\lint=0.002$ $\lsun$ (see text for discussion).  (d) Same as (a), except with no component from the Interstellar Radiation Field to the heating of the envelope.}
\end{figure}

Figure \ref{2dsed}a shows a representative model from the constraints given above.  The specific parameters used in this model are $T_s=3500$ K, $L_s=0.06$ $\lsun$, $L_d=0.02$ $\lsun$, $r_i=90$ AU, and $i=$49.5\mbox{$^{\circ}$}.  We calculate $\chi^2_1 = 114$.  The fit looks very similar to the 1-D case, but the quality of the fit is worse because of the higher number of free parameters in the 2-D models.  Furthermore, as discussed above, models with a larger envelope inner radius but also a larger inclination angle will produce about the same amount of mid-infrared emission.  Thus, Figure \ref{2dsed}b presents the same model as Figure \ref{2dsed}a, except with $i = 50\mbox{$^{\circ}$}$ and $r_i = 260$ AU.  The quality of the fit is about the same, as expected.  We consider this model to be less likely since it is only marginally consistent with the Belloche et al. (2002) constraint that the envelope inner radius  be less than $\sim$ 260 AU, but it can not be ruled out as a possibility.

As mentioned above, despite the fact that Figure \ref{2dcontours}d suggests that very small internal luminosities are allowed, \lint\ must be greater than $\sim 0.04$ \lsun.  Figure \ref{2dsed}c shows the best model with $\lint=0.002$ $\lsun$, the lowest internal luminosity considered in the grid of models.  The inclination for this model is $i=$49.0\mbox{$^{\circ}$}.  We calculate $\chi^2_1 = 117$, nearly identical to the best-fit model presented above.  With a slightly lower inclination than the above models, the line-of-sight passes more directly through the outflow cavity.  Thus, most of the mid-infrared emission from the internal source escapes directly through the envelope, leading to a much better fit at $3.6-8.0$ \um.  However, the fit is much worse at 24 and 70 \um, especially at 70 \um.  The net result of a significantly improved fit at $3.6-8.0$ \um\ but worsened fit at 24 and 70 \um\ is a nearly identicaly $\chi^2_1$.  Since the 70 \um\ observation is much less sensitive to the effects of geometry than those at shorter wavelengths, and also because the emission at this wavelength is dominated by envelope emission from reprocessing of the internal luminosity, we impose the requirement that the models must produce at least enough emission at 70 \um\ to match the observation.  This results in the constraint that the internal luminosity must be greater than $\sim 0.04$ \lsun.  The issue of fitting the SED at $3.6-8.0$ \um\ versus fitting the SED at 24 and 70 \um\ is discussed in greater detail in \S \ref{lint}.

\subsection{1-D vs. 2-D}\label{modelsummary}

Strong observational evidence exists for the presence of an outflow cavity in IRAM 04191.  The failure of our 1-D models, which are unable to include such a structure, supports this evidence.  However, simple 2-D models do not improve the fits over those obtained in 1-D.  Nevertheless, they are still important because they are more realistic, and they make predictions that can be tested with future observations.

Table \ref{bestmodels} presents our best 1-D and 2-D models of IRAM 04191.  The two key distinctions between them are the envelope inner radius and the shape of the SED in the $10 - 100$ \um\ range.  Smaller envelope inner radii are allowed for the 2-D models than the 1-D models due to the outflow cavities in the 2-D models allowing varying amounts of mid-infrared emission to escape.  High-resolution interferometer observations or \emph{Spitzer} Infrared Spectrograph (IRS) spectra could provide constraints on the inner regions of the envelope and perhaps allow us to rule out the 1-D models altogether (e.g., J\o rgensen et al. 2005), as well as provide insight into the extent of the envelope inner radius, a physical parameter that is intimately connected to the uncertain transition between the envelope and the disk.  IRS observations would also be very useful in examining the shape of the mid-infrared SED and comparing to the predictions of the various models, and such observations are currently planned.

Except for the internal luminosity, which is well-determined by our models (see \S \ref{lint}), the other parameters listed in Table \ref{bestmodels} should be treated with less confidence.  By assuming a conical outflow cavity with an opening half-angle of $50\mbox{$^{\circ}$}$ in the 2-D models, we are biased towards an inclination of approximately the same angle.  In reality, it is the difference between the opening half-angle of the outflow cavity and the inclination that the models constrain, rather than each parameter individually.  Furthermore, since the 1-D models were very insensitive to the stellar temperature and the 2-D models were only marginally sensitive to this parameter, we do not list any constraint on it in Table \ref{bestmodels}.  Both the stellar temperature and the inclination that best fit the data depend on the details of the 2-D structure.  This structure is likely significantly more complicated than the simple conical outflow cavity considered here, and the simplicity of these models may at least partially explain our inability to improve the fits over those obtained in 1-D.

Beyond the uncertain details of the 2-D structure, several other factors may be at work in preventing good fits to the $3.6-8.0$ \um\ SED.  For one, our dust models do not include several ice features known to exist in the mid-infrared (e.g., Boogert et al. 2004).  In a deeply embedded core such as this one, these ice features could have an important effect on the observed emission.  And, perhaps even more importantly, because our models are only capable of treating scattering as isotropic, we instead treat all of the opacity (scattering and absorption) as only absorption.  Including scattering, especially anisotropic scattering, in models where the line-of-sight intersects with an outflow cavity could increase the emission at the shortest wavelengths and thus provide a better fit to the data.

The difficulty in fitting the $3.6-8.0$ \um\ SED is not unique to IRAM 04191, but instead seems to be common for such Very Low Luminosity Objects (e.g., L1521F, Bourke et al. 2006).  While the longer-wavelength far-infrared and millimeter data are sensitive to the total luminosity of the source and mass of the envelope, the mid-infrared data is sensitive to both the details of the structure and geometry of the envelope and the exact shape of the extinction curve at these wavelengths.  Thus, for sources such as IRAM 04191, constructing SEDs from 3.6 \um\ to millimeter wavelengths allows us to investigate a wide range of parameters that would not be possible without the \emph{Spitzer} data at wavelengths less than 10 \um.  Future, more sophisticated modeling will concentrate on exploring these parameters in more detail.

\subsection{Constraining the Internal Luminosity}\label{lint}

The most important motivation for this modeling has been to provide an estimate of the internal luminosity of IRAM 04191-IRS, as an accurate determination of this quantity is essential in order to determine whether or not this object qualifies as a VeLLO.  The agreement between the 1-D and 2-D models on \lint\ suggests that they provide a reasonable estimate of this value even though they do not provide optimal fits to the observed $3.6-8.0$ \um\ SED.  This is true because most of the internal luminosity is reprocessed by the surrounding envelope and re-emitted at longer wavelengths.  In the absence of the Interstellar Radiation Field, the good fit between the models and the longer-wavelength data that dominates the total emission from the source would signify that our value for $L_{int}$ was correct.  However, with the ISRF present, it becomes more complicated.

Figure \ref{2dsed}d shows the same 2-D model as that shown in Figure \ref{2dsed}a, except with no Interstellar Radiation Field, and so the only source of heating is the internal source.  From the figure it is clear that most of the internal luminosity is emitted longward of 10 \um\ and that the emission longward of 100 \um\ is not fitted by this model, where the emission is dominated by cold dust heated by the ISRF.  To show that we predict the correct $L_{int}$, we define two new quantities: $L_{24-70}$, the luminosity predicted by the model between 24 and 70 \um, and $L_{IRAC}$, the luminosity predicted by the model between 3.6 and 8.0 \um.  For the model with no ISRF, we calculate $L_{24-70}/L_{IRAC} \sim 320$, while for the model with the ISRF included, we calculate $L_{24-70}/L_{IRAC} \sim 345$.  These two ratios agree to within less than 10\%, showing that the emission out to 70 \um\ is dominated by the internal luminosity.  Furthermore, these ratios illustrate that the emission between 24 and 70 \um\ contributes greater than two orders of magnitude more to the internal luminosity than the emission between 3.6 and 8.0 microns, and as a result, the exact fit to the $3.6-8.0$ \um\ SED has a negligible effect on $L_{int}$.  As the models presented in this work produce good fits in the $24-70$ \um\ range, they do provide reasonable estimates of the internal luminosity.

If the disk were observed edge-on, the internal luminosity could be underestimated.  We consider such a geometry to be unlikely based on the fact that the outflow shows a clear bipolar morphology with almost no overlap between blueshifted and redshifted emission, suggesting a moderate inclination for this object.  However, it must be ruled out entirely before any strong conclusions about the internal luminosity can be drawn.  The observed emission from $3.6-24$ \um\ arises from extincted light from the central object (protostar and disk, if present).  However, as seen by the comparison between the input spectrum and the model spectrum in Figure \ref{2dsed}d, the emission at 70 \um\ is dominated by envelope emission from reprocessing of the internal luminosity.  Even if an edge-on disk were obscuring the $3.6-24$ \um\ emission, any underestimate of the internal luminosity would be seen in the fit to the observation at 70 \um.

To determine an upper limit for \lint\ in such a case we increased the internal luminosity, ignoring the fit between $3.6-24$ \um, until the model exceeded the observation at 70 \um.  Over the full range of parameter space, no model with $\lint > 0.15$ \lsun\ was able to fit the data at 70 \um\ to within $3\sigma$, and no model with  $\lint > 0.1$ \lsun\ was able to fit the data at 70 \um\ to within $1\sigma$.  While we do not actually include an edge-on disk in these models, if a more luminous disk were present it would make the fits at 70 \um\ even worse by increasing the amount of emission from warm dust at this wavelength.  Thus, we conclude that our models predict an accurate value for the internal luminosity:  $\lint = 0.08 \pm 0.04$ $\lsun$.

\section{IRAM 04191-IRS: A Very Low Luminosity Object}\label{vello}

IRAM 04191-IRS is projected onto a dense core of gas and dust, as seen from both submillimeter and millimeter continuum emission (\andre\ et al. 1999; Belloche et al. 2002; Young et al. 2006; Wu et al. in preparation) and molecular line emission (\andre\ et al. 1999; Belloche et al. 2002; Takakuwa et al. 2003; Belloche \& \andre\ 2004; Lee et al. 2005).  Additionally, it is associated with both a CO outflow (\andre\ et al. 1999; also see Figure \ref{results2}), and extended, mid-infrared nebulosity that correlates with the edge of the blue-shifted outflow lobe.  Thus, this source is clearly embedded in a dense core, and with the conclusions from this work that $\lint = 0.08 \pm 0.04$ $\lsun$ and the definition given in \S \ref{intro} that a VeLLO is an object embedded within a dense core with an internal luminosity $\lint \leq 0.1$ \lsun, we conclude that IRAM 04191-IRS qualifies as a VeLLO.

IRAM 04191-IRS was known to exist prior to the \emph{Spitzer} observations, based primarily on the presence of its strong outflow.  Placing its internal luminosity and outflow parameters given in Lee et al. (2002) and \andre\ et al. (1999) (Mass $M = 3 \times 10^{-2}$ $\msun$; Dynamical time $t_d = 0.8 \times 10^4$ years; Force $F_{obs} = 1.5 \times 10^{-5}$ \msun\ km s\mbox{$^{-1}$} yr\mbox{$^{-1}$}) on the plots of the relations between these quantities given in Wu et al. (2004) suggests it is consistent with the relations found for higher luminosity ($\lint \geq 1$ $\lsun$) sources.  Thus, at least in the properties of its outflow, IRAM 04191-IRS appears to simply be a lower luminosity version of a ``normal'' protostar.

Taking the average mass accretion rate implied by the outflow driven by IRAM 04191-IRS and the dynamical time of this outflow ($\langle$$\dot{M}_{acc}$$\rangle$ $\sim 5 \ee{-6}$ \msun\ yr$^{-1}$ and $t_d \sim 10^4$ years), \andre\ et al. (1999) calculate that a protostellar mass of 0.05 \msun will have accreted over the lifetime of the outflow.  Accretion at this rate onto an object with a mass of $0.05$ \msun\ and a radius of 3 \rsun\ would give rise to an accretion luminosity, $L_{acc} \sim \frac{GM\dot{M}_{acc}}{R}$, of $\sim 2$ \lsun.  This is more than an order of magnitude higher than $\lint \sim 0.08$ $\lsun$ from this work, and if the radius were much smaller the discrepancy between the internal and accretion luminosities would be even worse.  In order to resolve this discrepancy, the product of the mass and mass accretion rate must be much lower than assumed in this calculation.  A plausible way to accomplish this is to invoke non-steady accretion:  the current mass accretion rate must be much lower than the average rate over the lifetime of the outflow.  In other words, $\dot{M}_{acc}(t)$ $\neq$ $\langle$$\dot{M}_{acc}$$\rangle$.

The standard model of star formation (Shu, Adams, \& Lizano 1987) predicts a mass accretion rate that is constant with time.  However, non-steady accretion rates have previously been invoked to explain FU Orionis objects (Hartmann \& Kenyon 1985) and the low luminosities of many embedded objects in Taurus (Kenyon \& Hartmann 1995).  A possible mechanism for such varying mass accretion rates is that matter accretes from the envelope onto a circumstellar disk at a more or less uniform rate, but does not immediately accrete from the disk onto the protostar itself.  Instead, it is stored in the disk for a period of time before accreting onto the star in a short-lived burst (e.g., Vorobyov \& Basu 2005).  In this picture, IRAM 04191-IRS is perhaps seen in a quiescent phase of a cycle of episodic accretion, with a mass accretion rate much lower than in the burst phases.

In addition to IRAM 04191-IRS, two other VeLLOs have been studied in detail to date (L1014-IRS, Young et al. 2004a; L1521F-IRS, Bourke et al. 2006).  All three objects have similar internal luminosities: $\lint \sim 0.08$ \lsun\ for IRAM 04191-IRS, $\lint \sim 0.06$ \lsun\ for L1521F-IRS, and $\lint \sim 0.09$ \lsun\ for L1014-IRS.  Unlike IRAM 04191-IRS, the other two VeLLOs do not drive strong outflows and thus were not previously known to exist.  L1014-IRS is associated with a very compact, weak outflow (Bourke et al. 2005) not detected in single-dish observations (Crapsi et al. 2005b).  Placing the parameters of its outflow given in Bourke et al. on the plots of Wu et al. (2004) in the same manner as IRAM 04191-IRS shows that L1014-IRS has an outflow much less consistent with those of higher luminosity sources.  Instead, the outflow appears to be less massive and less luminous for the given internal luminosity of the source than would be inferred by extrapolating the relations found for higher luminosity sources down to the luminosity of L1014-IRS.  The other VeLLO studied in detail to date, L1521F-IRS, shows possible evidence for an outflow in single dish observations (Bourke et al. 2006), but it is neither extended nor well-defined.  Future interferometer observations are required to examine this source in more detail.

Any attempt to invoke non-steady accretion rates through a cycle of episodic accretion in order to explain the very low luminosities of these objects must be able to account for the different outflow properties between these three VeLLOs and, as implied through this, the different mass accretion histories.  Even though all three VeLLOs are embedded in envelopes with greater than 1 \msun\ of material (Young et al. 2004; Bourke et al. 2006), the very low combination of mass and mass accretion rates required to explain the luminosities of these objects makes their final masses quite uncertain.  It is possible that they may in fact be embedded proto-brown dwarfs, but further investigation is required before any strong conclusions can be drawn.  Future work will concentrate on assembling a more complete sample of VeLLOs and refining current modeling techniques in order to assess the validity of the picture of episodic accretion and determine the ultimate fate of these objects.

\section{Conclusions}\label{conclusions}

For the first time, we report detections of the Class 0 protostellar source IRAM 04191+1522 at wavelengths shortward of 60 \um.  There is strong evidence that an outflow cavity is present in the envelope surrounding IRAM 04191-IRS.  We find $L_{int} = 0.08 \pm 0.04$ $\lsun$, thus classifying IRAM 04191-IRS as a member of a new class of Very Low Luminosity Objects.  We also make predictions on the inner radius of the envelope and the shape of the SED from $10 - 100$ \um\ that can be tested with future observations, such as interferometric observations and IRS spectra, to distinguish between the various models presented.  Furthermore, we note that such observations, especially high-resolution submillimeter and millimeter interferometric observations, are necessary in order to confirm the apparent small size of the circumstellar disk surrounding IRAM 04191-IRS.  It the disk does turn out to be as small as suggested in previous work and as assumed in this work, it is unclear whether such a small size is due to being in a very early stage of formation or is perhaps instead connected to the very low luminosity of this object.  In either case, IRAM 04191 could become a particularly interesting source for the study of the earliest stages of disk formation and evolution.

Future work on this object will concentrate on more detailed 2-D modeling in order to better constrain the structure and geometry of the outflow cavity.  It will also concentrate on a more correct treatment of the radiative transfer that includes scattering, and with this, on reproducing the extended emission seen in the images between 3 and 8 \um.

The internal luminosity of IRAM 04191-IRS is similar to the luminosities of recent \emph{Spitzer} discoveries in L1014 and L1521F, but it drives a much more powerful outflow than either of the sources in these other cores.  The outflow driven by IRAM 04191-IRS implies a non-steady accretion rate that can perhaps be explained by a cycle of episodic accretion, but the differences between the outflows driven by this source and by the other two VeLLOs studied to date are not yet understood.  Future work is necessary in order to build a more complete sample of VeLLOs and to examine their implications for the current understanding of both low-mass star and brown dwarf formation.

We would like to thank Chin-Fei Lee for allowing us to present his BIMA data in this work, Jes J\o rgensen for his IDL scripts to display three-color images, and the Lorentz Center in Leiden for hosting several meetings that contributed to this paper.  We express our gratitude to the anonymous referee for insightful and thoughtful comments that have greatly improved this paper.  This publication makes use of the Protostars Webpage hosted by the Dublin Institute for Advanced Studies, as well as data products from the Two Micron All Sky Survey, which is a joint project of the University of Massachusetts and the Infrared Processing and Analysis Center/California Institute of Technology, funded by the National Aeronautics and Space Administration and the National Science Foundation.  Support for this work, part of the Spitzer Legacy Science Program, was provided by NASA through contracts 1224608 and 1230782 issued by the Jet Propulsion Laboratory, California Institute of Technology, under NASA contract 1407.  Support was also provided by NASA Origins grant NNG04GG24G.

\clearpage
\begin{table}[t]
\caption{\label{iramphot}Photometry of IRAM 04191+1522}
\begin{tabular}{lcccc}
\hline
$\lambda$ & $S_{\nu}(\lambda )$ & $\sigma$ & Aperture & Reference\\
(\um ) & (mJy) & (mJy) & (arcsec) \\
\hline
1.25 & $<$ 0.767 & - & - & 1 \\
1.65 & $<$ 0.980 & - & - & 1 \\
2.17 & $<$ 1.330 & - & - & 1 \\
2.2 & $<$ 0.100 & - & - & 3 \\
3.6 & 0.19  & 0.03 & 1.7$^a$ & 2\\
4.5 & 0.85  & 0.13 & 1.7$^a$ & 2\\
5.8 & 0.92  & 0.14 & 1.9$^a$ & 2\\
7.8 & $<$ 10.0 & - & - & 3\\
8.0 & 0.57  & 0.09 & 2.0$^a$ & 2\\
15 & $<$ 10.0 & - & - & 3\\
24 & 13.7  & 2.1 & 6.0$^a$ & 2\\
25 & $<$ 100.0 & - & - & 3 \\
60 & 500  & 100 & 60 & 3\\
70 & 600 & 120 & 50 & 2\\
90 & 800  & 160 & 60 & 3\\
160 & 6000  & 1200 & 60 & 3\\
200 & 10000  & 2000 & 60 & 3\\
350 & 5000 & 800 & 40 & 4\\
450 & 10000  & 2000 & 60 & 3\\
850 & 2500  & 500 & 60 & 3\\
1300 & 650  & 65 & 60 & 3\\
\hline
\end{tabular}\\
$^a$FWHM of \emph{Spitzer} point-spread profile\\
1 - Two Micron All Sky Survey\\
2 - New \emph{Spitzer} observations\\
3 - \andre\ et al. 1999\\
4 - Wu et al. (in preparation)
\end{table}
\begin{table}[t]
\caption{\label{irasphot}Photometry of IRAS 04191+1523}
\begin{tabular}{lccccc}
\hline
Component & $\lambda$ & $S_{\nu}(\lambda )$ & $\sigma$ & Aperture & Reference\\
 & (\um ) & (mJy) & (mJy) & (arcsec) & \\
\hline
IRAS 04191+1523-A & 1.25 & 0.3 & 0.03 & 2.5$^a$ & 1\\
 & 1.65 & 1.7 & 0.2 & 2.5$^a$ & 1\\
 & 2.17 & 8.3 & 0.9 & 2.5$^a$ & 1\\
 & 3.6 & 47 & 5 & 1.7$^b$ & 2\\
 & 4.5 & 95 & 10 & 1.7$^b$ & 2\\
 & 5.8 & 130 & 10 & 1.9$^b$ & 2\\
 & 8.0 & 170 & 20 & 2.0$^b$ & 2\\ \\
IRAS 04191+1523-B & 1.25 & 0.06 & 0.01 & 2.5$^a$ & 1\\
 & 1.65 & 0.40 & 0.04 & 2.5$^a$ & 1\\
 & 2.17 & 0.9 & 0.1 & 2.5$^a$ & 1\\
 & 3.6 & 4.0 & 0.4 & 1.7$^b$ & 2\\
 & 4.5 & 6.7 & 0.7 & 1.7$^b$ & 2\\
 & 5.8 & 6.5 & 0.7 & 1.9$^b$ & 2\\
 & 8.0 & 7.3 & 0.7 & 2.0$^b$ & 2\\ \\
IRAS 04191+1523 & 24 & 780 & 80 & 6.0$^b$ & 2\\
 & 70 & 3600 & 900 & 50 & 2\\
 & 450 & 1400 & 800 & 40 & 3\\
 & 850 & 600 & 100 & 40 & 3\\
 & 1300 & 400 & -$^c$ & 60 & 4\\
\hline
\end{tabular}\\
$^a$FWHM of Two Micron All Sky Survey point-spread profile\\
$^b$FWHM of \emph{Spitzer} point-spread profile\\
$^c$No flux uncertainty was listed\\
1 - Two Micron All Sky Survey\\
2 - New \emph{Spitzer} Observations\\
3 - Young et al. (2006)\\
4 - Motte \& \andre\ (2001)\\
\end{table}
\begin{table}[t]
\caption{\label{bestmodels}Best-fit Models of IRAM 04191+1522$^a$}
\begin{tabular}{lll}
\hline
Parameter & 1-D & 2-D \\
\hline
Internal Luminosity $(L_{int})$ & $0.08 \pm 0.04$ $\lsun$ & $0.08 \pm 0.04$ $\lsun$ \\
Bolometric Luminosity $(L_{bol})$ & $0.28 \pm 0.06$ $\lsun$ & $0.28 \pm 0.06$ $\lsun$\\
Envelope Inner Radius $(r_i)$ & $150 \leq r_i \leq 260$ AU & $90 \leq r_i \leq 260$ AU \\
Inclination $(i)$ & - & $\sim 50\mbox{$^{\circ}$}$ $^b$ \\
\hline
\end{tabular}\\
$^a$The stellar temperature is not included in this table because, as discussed in the text, these models do not constrain this parameter.\\
$^b$By assuming a conical outflow cavity with an opening half-angle of $50\mbox{$^{\circ}$}$ in the 2-D models, we are biased towards an inclination of approximately the same angle.  Other outflow cavity sizes and geometries would likely result in other inclinations.\\
\end{table}

\clearpage

\end{document}